\documentclass[fleqn,10pt]{wlscirep}
\usepackage{color}
\usepackage[english]{babel}
\usepackage{lineno}

\usepackage{graphics}
\usepackage{graphicx}
\usepackage{amsmath}
\usepackage{epsfig}
\usepackage{fancyhdr}
\usepackage{wrapfig}  
\usepackage{amsmath,amssymb}
\usepackage{tabularx}
\usepackage{cite} 
\usepackage{url} 
\usepackage{multicol}
\usepackage{multirow}

\begin{document}

\title{Bose--Einstein Condensation in a Plasmonic Lattice}

\author[1,$\dagger$]{Tommi K. Hakala}
\author[1,$\dagger$]{Antti J. Moilanen}
\author[1]{Aaro I. V\"{a}kev\"{a}inen}
\author[1]{Rui Guo}
\author[1]{Jani-Petri Martikainen}
\author[1]{Konstantinos S. Daskalakis}
\author[1]{Heikki T. Rekola}
\author[1]{Aleksi Julku}
\author[1,*]{P\"{a}ivi T\"{o}rm\"{a}}
\affil[1]{COMP
Centre of Excellence, Department of Applied Physics, Aalto University School of Science, FI-00076 Aalto, Finland}

\affil[*]{paivi.torma@aalto.fi}
\affil[$\dagger$]{These authors contributed equally to this work.}

\begin{abstract}
Bose--Einstein condensation is a remarkable manifestation of quantum statistics and macroscopic quantum coherence. Superconductivity and superfluidity have their origin in Bose--Einstein condensation. Ultracold quantum gases have provided condensates close to the original ideas of Bose and Einstein, while condensation of polaritons and magnons have introduced novel concepts of non-equilibrium condensation. Here, we demonstrate a Bose--Einstein condensate (BEC) of surface plasmon polaritons in lattice modes of a metal nanoparticle array. Interaction of the nanoscale-confined surface plasmons with a room-temperature bath of dye molecules enables thermalization and condensation in picoseconds. The ultrafast thermalization and condensation dynamics are revealed by an experiment that exploits thermalization under propagation and the open cavity character of the system. A crossover from BEC to usual lasing is realized by tailoring the band structure. This new condensate of surface plasmon lattice excitations has promise for future technologies due to its ultrafast, room-temperature and on-chip nature. 

\end{abstract}

\flushbottom
\maketitle

\thispagestyle{empty}

Bosonic quantum statistics imply that below a certain critical temperature or above a critical density the occupation of excited states is strictly limited, and consequently, a macroscopic population of bosons accumulates on the ground state~\cite{Griffin1995}. This phenomenon is known as Bose--Einstein condensation (BEC). Superconductivity of metals and high-temperature superconducting materials are understood as BEC of Cooper pairs~\cite{LeeRMP2005,Zwerger2012}.
The BEC phenomenon is central in superfluidity of helium although the condensate constitutes a small fraction of the particles~\cite{Volovik2003}. Textbook Bose--Einstein condensates with large condensate fractions and weak interactions were created with ultracold alkali atoms~\cite{CornellBEC1995,KetterleBEC1995,BradleyHuletBEC1995}, 
and the fundamental connection between the superfluidity of Cooper pairs and the Bose--Einstein condensation was confirmed by experiments with ultracold Fermi gases~\cite{Zwerger2012}. While all these condensates allow essentially equilibrium description, as was the original one by Bose and Einstein, the phenomenology has expanded to non-equilibrium systems~\cite{imamoglu_nonequilibrium_1996,deng_exciton-polariton_2010,carusotto_quantum_2013,Altman2013,byrnes_exciton-polariton_2014}. Hybrid particles of semiconductor excitons and cavity photons, called exciton-polaritons, have shown condensation and interaction effects~\cite{deng_condensation_2002,kasprzak_boseeinstein_2006,balili_bose-einstein_2007,baumberg_2008,Amo2009,daskalakis_nonlinear_2014,plumhof_room-temperature_2014}, creating coherent light output that deviates from usual laser light. Magnons, that is, spin-wave excitations in magnetic materials~\cite{demokritov_boseeinstein_2006,giamarchi_boseeinstein_2008}, and photons in microcavities~\cite{klaers_bose-einstein_2010,marelic_spatiotemporal_2016} form condensates as well. 
The most technologically groundbreaking manifestation of macroscopic population due to bosonic statistics has so far been laser light, which is a highly non-equilibrium state not thermalized to a temperature of any reservoir. As the BEC phenomenon has been observed only in a limited number of systems, new ones are needed for pushing the time, temperature and spatial scales where a BEC can exist, as well as for opening viable routes to technological applications of BEC.

Here we report the observation of BEC for bosonic quasiparticles that have not previously been shown to condense: excitations of surface lattice resonance (SLR) modes~\cite{zou_silver_2004,garcia_de_abajo_colloquium:_2007,auguie_collective_2008,rodriguez_thermalization_2013,Martikainen2014} in a metal nanoparticle array. SLRs originate from coupling between localized surface plasmon resonances of the nanoparticles and diffracted orders of an optical field incident on the periodic structure. The quasiparticle is mostly of photonic nature but also contains a matter part, namely electron plasma oscillations in the nanoparticles. This is different from the excitonic matter part in polariton condensates~\cite{deng_exciton-polariton_2010,carusotto_quantum_2013,byrnes_exciton-polariton_2014,deng_condensation_2002,kasprzak_boseeinstein_2006,balili_bose-einstein_2007,baumberg_2008,Amo2009,daskalakis_nonlinear_2014,plumhof_room-temperature_2014}, and distinguishes our system from the purely photonic condensates~\cite{klaers_bose-einstein_2010,marelic_spatiotemporal_2016}. The SLR excitations are essentially non-interacting, therefore we use weak coupling interaction with a molecular bath~\cite{klaers_bose-einstein_2010,schmitt_kinetics_2015,marelic_spatiotemporal_2016} to achieve thermalization. 
Due to the plasma oscillation component, losses fundamentally limit~\cite{khurgin_ultimate_2015} the lifetimes of plasmonic systems~\cite{Maier2001,Novotny2011} typically to 10 fs -- 1 ps. 
To achieve a condensate with Bose--Einstein statistics at room temperature, we have predicted~\cite{Martikainen2014, Julk15, Moil16} that thermalization timescales need to be 2--3 orders of magnitude shorter than in photon condensates~\cite{klaers_bose-einstein_2010} and similar to the fastest dynamics in polariton condensates~\cite{baumberg_2008,daskalakis_nonlinear_2014,plumhof_room-temperature_2014}. Lasing has been observed in plasmonic lattices in the weak coupling~\cite{zhou_lasing_2013,hakala_lasing_2017} and strong coupling (polariton)~\cite{ramezani_lasing2017} regimes. To compare SLR lasing and BEC, we realize an experimental setting that provides direct access to the evolution of thermalization and condensate formation in the sub-picosecond scale.

\begin{figure}[t!]
  \centering
    \includegraphics[width=0.8\textwidth]{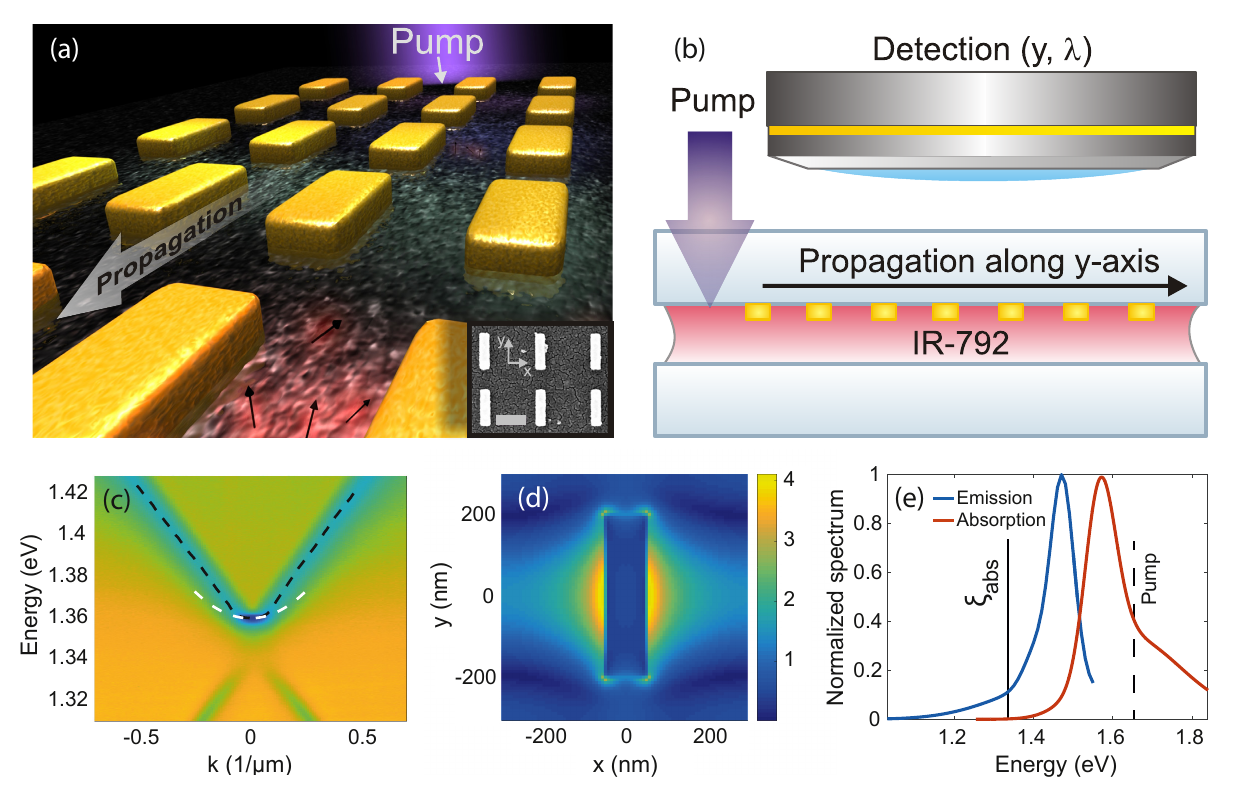}
   \caption{\footnotesize{\textbf{Gold nanoparticle array and energy dispersion.} a) Schematic of the system: a gold nanoparticle array combined with IR-792 dye molecules in solution, the inset shows scanning electron microscope (SEM) image of a part of the array (the scale bar is 300 nm); the actual array size is 100 x 300 $\mathrm{\mu} \mathrm{m}^2$. b) Schematic of the experimental scheme. c) Dispersion $E(k_y)$ as function of the in-plane momentum $k_y$ (the polarization of the field in this mode is along the $x$-direction) of a bare array without molecules obtained by a white light transmission measurement, showing a band edge whose position in energy can be tuned by the inter-particle distance. Maximum of the dispersion is marked by a blue dashed line and a parabola for extracting an effective mass by a white dashed line. d) The absolute value of the electric field in the array plane relative to the incident field at pump pulse maximum, $|{\mathbf E}|/|{\mathbf E}_0|$, in one unit cell, calculated by the finite-difference time-domain method. e) Emission and absorption profiles of the molecule: the solid line indicates the point in energy ($\xi_{\mathrm{abs}}$) where absorption is effectively zero (i.e., small compared to the losses), and the dashed line shows the pump energy. The emission maximum is at 1.46 eV.
}}
   \label{figure1}
\end{figure}

\section*{The system}

Our systems constitute of periodic arrays of gold nanoparticles on a glass substrate with inter-particle distances of 580--610 nm and array dimensions of 100 $\times$ 300 $( \mathrm{\mu m} )^2 $, see Figs.~\ref{figure1}a-b, and Methods. Figure~\ref{figure1}c shows the measured dispersion of the SLR mode. Due to coupling between the counter-propagating modes at ${\bf k}=0$ ($\Gamma$-point), a band gap opening typical for a grating is observed with a band edge in the upper dispersion branch. The curvature of the band bottom corresponds to an effective mass $m_{eff} \sim 10^{-37}$ kg ($10^{-7}$ electron masses), obtained by fitting to a parabola, Fig.~\ref{figure1}c. The linewidth of the SLR mode near the band edge is 5.5 meV (3.8 nm) and the lifetime is 120 fs. 

We use 50 mMol concentration molecule solution for which weak coupling between the SLR modes and the molecules is expected~\cite{torma_strong_2015}. Figure 1e shows the absorption and emission profiles of the molecule (IR-792 perchlorate $\mathrm{C}_{42}\mathrm{H}_{49}\mathrm{ClN}_{2}\mathrm{O}_{4}\mathrm{S}$), and the point in energy ($\xi_{\mathrm{abs}}$), where absorption rate is effectively zero (i.e., small compared to the loss rate), is marked as a solid line. We excite the molecules with a 100 fs laser pulse and measure the far-field emission.  

\begin{figure}[t!]
  \centering
    \includegraphics[width=1\textwidth]{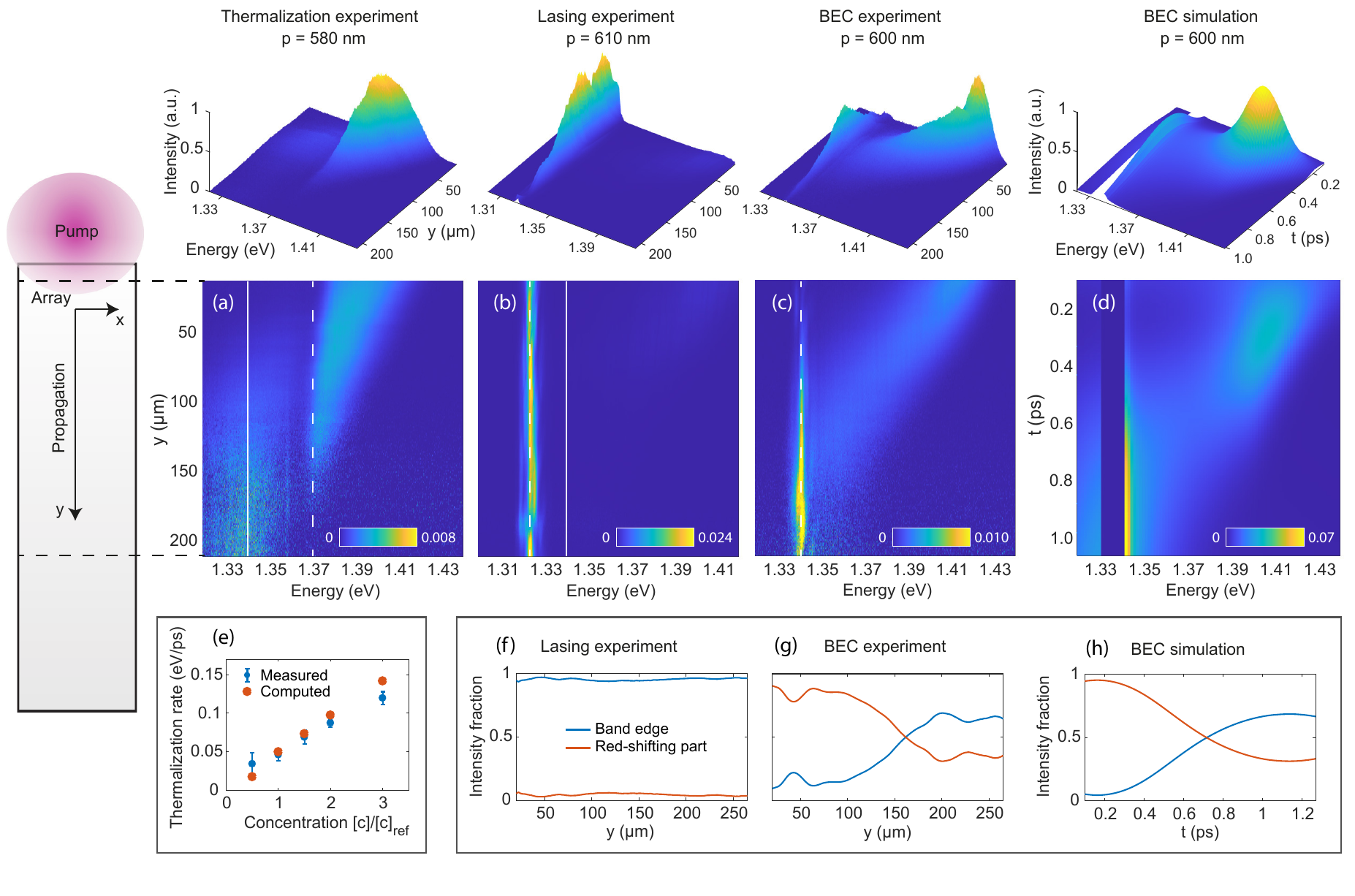}
   \caption{\footnotesize{\textbf{Spectral and spatial evolution of the sample luminescence: crossover from thermalization to lasing to BEC.} Photon part of the SLR population constantly leaks out and we can measure its spectral evolution as a function of position in $y$-direction. The insets above a)-c) show the {\it raw} intensity data as 3D spectra for periodicities 580, 610 and 600 nm, respectively. a)-c) show the same data so that the intensity of each row is normalized by the summed total intensity at that row. The band edge is marked with a dashed line in each case, and $\xi_{\mathrm{abs}}$ with a solid line in a)-b). In c) the band edge and $\xi_{\mathrm{abs}}$ are at the same energy (dashed line). f)-g) show how the relative intensities of the band edge and the rest of the population at higher energies evolve along the array in the lasing and BEC cases, respectively. 
In f) and g) the intensities of the band edge (1.32--1.33 eV and 1.34--1.35 eV) and the red-shifting part of the population (1.33--1.41 eV and 1.35--1.43 eV) are normalized by the summed total intensity over the energies 1.32--1.41 eV and 1.34--1.43 eV, respectively. d) and h) display the simulation results of the dynamics in time-domain, for parameters corresponding to the BEC case. e) Thermalization rate as a function of concentration. Measured data shows the mean value from several measurements for each concentration ($[c]_\mathrm{ref}$ = 50 mMol), with error bars showing the standard deviation. For simulation, the $[c]_\mathrm{ref}$ corresponds to the number of molecules $N$ used for calculating the results in d) and h). For details of the thermalization rate measurement, see Methods. Colour scales in a)-c) are saturated for better visibility of interesting features, the maximum intensity values are a) 0.0096, b) 0.0286 and c) 0.0175. Note that in a)-c) the data from the edge of the array close to the pump spot has not been shown.
}}
   \label{figure2}
\end{figure}

The SLR modes are confined via the near-field character of the nanoparticle plasmonic resonances as well as by interference effects due to the periodic structure. The majority of the field intensity lies close (within a few hundreds of nanometers) to the array surface. Thus the array is an open system that can be easily imaged and pumped from the top; changing the properties of the modes, such as quality factor, does not affect the direct access for pumping the molecules. We utilize this to monitor the spatial evolution of the excitations, see Figs.~\ref{figure1}a-b. 

\section*{Observation of Bose--Einstein Condensation (BEC)}

We pump the dye molecules at the edge of the array and record the spectral evolution of SLR excitations propagating along the array, as sketched in Figs.~\ref{figure1}a-b. We investigate three samples that are otherwise identical except that the inter-particle distance is varied to tune the energy of the SLR band edge a) above, b) below, and c) matching
$\xi_{\mathrm{abs}}$. For intermediate cases, see Extended Data Fig.~1.

In Fig.~\ref{figure2}a, we observe first emission around the emission maximum of the molecule and then a gradual decrease in energy (red-shift) when the population propagates along the array. 
When a photon is absorbed to one of the rovibrational levels of the molecule excited state, the excitation is likely to relax to some of the lower levels before a photon is emitted. Such recurrent emission and absorption cycles provide the red-shift. In the case of Fig.~\ref{figure2}a, the band edge energy (dashed line) is well above $\xi_{\mathrm{abs}}$ (solid line). The molecules offer a bridge over the band gap, as photons are absorbed from the energies of the upper dispersion branch and emitted to the lower one. Eventually the red shift saturates to a broad spectral range at 1.34~eV, where the absorption has ceased (this is how we define $\xi_{\mathrm{abs}} \approx$ 1.34~eV).  

The occupation of the dye molecule rovibrational states follow Maxwell--Boltzmann statistics at room temperature due the contact of the dye molecules with solvent molecules, thus the molecules provide a room-temperature thermal bath for the SLR excitations to exchange energy. Excitation by the pump may, however, drive the molecules out of equilibrium. Note that in our system, the SLR excitations propagate away from the initial excitation spot and there are always new molecules in thermal equilibrium available.

The spatial evolution over the array can be mapped to time via the group velocity of the SLR mode everywhere except a few meV range of energy near the band edge, where the group velocity is vanishingly small. Thus we can extract the thermalization time (or rate) from the slope of the propagating population. Absorption probabilities can be increased by number $N$ of the molecules, which leads to approximately linear dependence of the thermalization rate on $N$~\cite{schmitt_kinetics_2015}. We varied the molecular concentration and measured the thermalization rate, giving the expected linear dependence, Fig.~\ref{figure2}e. In addition, the thermalization may be accelerated by the field hot spots near the nanoparticles (Fig.~\ref{figure1}d), as Purcell enhancement factors of the order of hundreds have been reported for similar systems~\cite{zhou_lasing_2013,dridi_model_2013}.

In Fig.~\ref{figure2}b, the system shows lasing
at the band edge similarly as reported before in almost identical systems but pumping over the whole array, see~\cite{zhou_lasing_2013,hakala_lasing_2017,ramezani_lasing2017} and references therein. 
Here the band edge is lower in energy than $\xi_{\mathrm{abs}}$, and thus absorption-emission cycles causing the red-shift are suppressed near the band edge.
The linewidth of the emission peak is around 1.5 nm, which is narrower than the natural linewidth of the SLR modes (3.8 nm). 
This, together with a non-linear pump dependency with a clear threshold, is characteristic for lasing (see Extended Data Fig.~2).

The situation changes drastically when the band edge
is matched with $\xi_{\mathrm{abs}}$, Fig.~\ref{figure2}c. 
The emission at the band edge energy is negligible at the beginning of the array. Now the array periodicity is chosen so that absorption-emission cycles are possible until the band edge energy and thermalization may take place. A macroscopic occupation emerges when the thermalizating population reaches the band edge.
In the following, we show that this phenomenon fulfills the main characteristics of a BEC: population following Bose--Einstein distribution with macroscopic occupation of the ground state and a thermalized tail, sharp increase in temporal coherence as well as build up of spatial coherence.

The difference between lasing and BEC is clearly shown in the evolution of the relative intensity at the band edge, and that of the rest of the population at higher energies, see Fig.~\ref{figure2}f-g. In the lasing case, a lasing peak is generated at the pump spot (for spectra collected from the part of the array that is under the pump spot,
see Extended Data Fig.~3) and then spreads through the array. In contrast, in the BEC case the population at the band edge is initially vanishing, and the thermalizing population is crucial for the emergence of the condensate peak. Thus we have a {\it direct observation that the population accumulated to the ground state has emerged via thermalization}. At around 150 $\mathrm{\mu m}$ the band-edge population becomes dominant, and after 200 $\mathrm{\mu m}$ the relative intensities of the band edge and the rest of the population stabilize. This is the area where we observe a room-temperature Bose--Einstein distribution. With the thermalization mechanism we use~\cite{klaers_bose-einstein_2010}, quasi-equilibrium can be reached if multiple emission-absorption cycles take place~\cite{Kirt13,Carusotto15}. In our lasing case, absorption is suppressed, and therefore the thermal tail is vanishing (Fig.~\ref{figure2}b,f and Extended Data Fig.~3). In the BEC case, the \textit{SLR excitations that form the condensate around 150 $\mu$m and thereafter are precisely those that have undergone several absorption-emission cycles} (Fig.~\ref{figure2}c,d,g,h). Therefore a Bose-Einstein (BE) distribution in our system has its origin in thermalization and condensation phenomena. This is important in order to distinguish from mere coexistence of a sharp peak and and a Boltzmann tail as observed for photonic lasing in semiconductor microcavities~\cite{Bajoni07}.

\begin{figure}[ht!]
  \centering
    \includegraphics[width=1.0\textwidth]{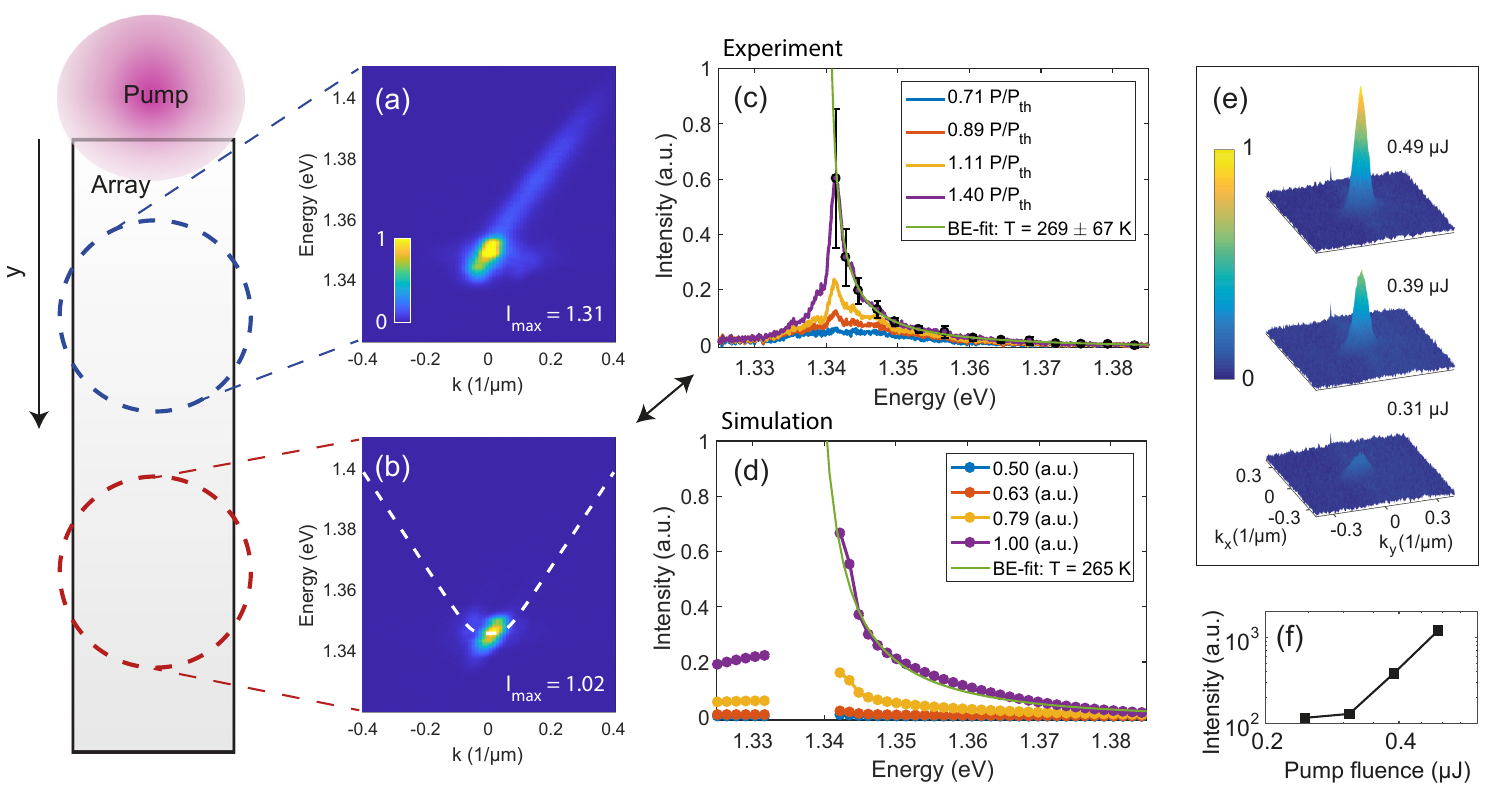}
   \caption{\footnotesize{\textbf{Momentum and energy distributions.} Spatially resolved $k$-space images. The array was pumped from one end, and the measurement corresponding to a) was done at the location indicated by the blue dashed circle in schematic on the left, while those in b), c), e) and f) correspond to data collected at the position marked by the red dashed circle. The dashed line in b) marks the fit to dispersion of SLR modes without molecules, Fig.~\ref{figure1}c. Data in a) shows that initially the thermalizing population follows the SLR dispersion while in b) the population has condensed to the band edge. Note that the asymmetry with respect to the pumping spot makes only one of the dispersion arms visible in the spectra. Colour scales in a)-b) are normalized to the same absolute intensity value, and saturated for visibility of the thermalizing population. c) The lines represent the measured raw population distributions at the $y$-position of the array denoted by the red dashed circle, and d) the circles connected with lines show the computed distributions at corresponding time. Bose--Einstein (BE) distribution fits are shown with solid green lines. The ratios between the consecutive pump fluencies are exactly the same in the experiment and in the simulations. The experimental data fits excellently to a BE distribution close to room temperature, $T=269 \pm 67$ K (95\% confidence bounds). The error bars show standard deviation for ten measurements. In contrast, fit to the Maxwell--Boltzmann (MB) distribution is poor (Extended Data Fig.~6b). e) 2D $k$-space measurement at the three highest pump fluencies, corresponding to the ones used in c), shows emergence of the condensate and its confinement in both $k_x$ and $k_y$. f) Intensity at the band edge as function of the pump fluence, showing threshold behaviour.
  }}
   \label{figure3}
\end{figure}

To further confirm that the observed luminescence originates from the SLR modes, we performed the measurements also in the momentum ($k$) space. As shown in Fig.~\ref{figure3}a-b, the population indeed follows the SLR dispersion down in energy until the band edge. It follows one dispersion arm in Fig.~\ref{figure3}a because, in the beginning of the array, only one SLR propagation direction is possible. 

{\bfseries Transition as function of pump fluence}
Transition to BEC should occur in a non-linear manner at a critical density or temperature. Fig.~\ref{figure3}c shows the measured population distributions for the BEC case at different pump fluencies. 
Using the relation between the critical density and the thermal de Broglie wavelength $\lambda_T = \sqrt{2\pi \hbar^2/(m_{eff} k_B T)}\simeq 10$ $\mathrm{\mu m}$ for a finite two-dimensional (2D) system, namely $n_c = 2/\lambda_T^2 \ln (L/\lambda_T)$~\cite{ketterle_bose-einstein_1996,deng_exciton-polariton_2010}, we obtain critical density of the order $10^{10}/$m$^2$ at $T=300$ K ($\hbar$ is the Planck's constant and $k_B$ the Boltzmann constant). Finite systems with 2D linear or 1D parabolic dispersions give the same order of magnitude, and 1D linear an order of magnitude smaller. For the system size $L$, we used the measured coherence length of the SLR mode in the absence of molecules, $L = 2\pi /\Delta k =45$ $\mathrm{\mu m}$. It was obtained from $\Delta k$ of the dispersion and confirmed by a double slit experiment, see Extended Data Fig.~4. Accurate determination of the density of photons at our sample is difficult, but the intensity of emitted photons gives it a lower bound. The lower bound is found to be higher, $10^{11}/$m$^2$, than the critical density estimate (see Supplementary Information).

At small pump fluencies, the measured spectrum in Fig.~\ref{figure3}c follows Maxwell--Boltzmann distribution (see Methods), but a peak emerges near the band edge when pump fluence is increased. After exceeding the threshold pump fluence, a spectrally sharp peak appears in the band edge with a long thermalized tail. The observed linewidth (1.5 nm) is narrower than the natural linewidth of the SLR modes, see also Extended Data Fig.~5. A decrease of the linewidth by around a factor of two is a direct indication of increase of the temporal coherence. Remarkably, the BE distribution fits excellently to the measured spectra in Fig.~\ref{figure3}c, with $T=269 \pm 67$ K, close to room temperature. Fits to the BE distribution for lower pump fluencies are shown in Extended Data Figure 6a. The distribution is an average over ten measurements, for details see Methods. In contrast, for the sample exhibiting lasing, the thermalized tail is vanishing and the fit to the BE distribution gives a temperature of only 25 K, see Extended Data Fig.~2.

{\bfseries Comparison to theory} The interpretation of our experimental observations is supported by a rate-equation approach~\cite{Kirt13}, see Methods. 
The computational spectral evolution in time-domain maps into the spatially measured spectra via group velocity of the SLR excitations. The experimentally observed evolution of the population, Fig.~\ref{figure2}c and Fig.~\ref{figure2}g, is in a good agreement with the computational result shown in Fig.~\ref{figure2}d and Fig.~\ref{figure2}h. For comparison between experiments in Fig.~\ref{figure2}a-b and the theory, see Extended Data Fig.~7. The observed thermalization times are typically 0.5--1 ps. A pump-probe experiment further confirms the picosecond time-scale of the dynamics (Extended Data Fig.~8).
The pump dependence of the population distribution (Fig.~\ref{figure3}c-d) and the thermalization time dependence on concentration (Fig.~\ref{figure2}e) show good agreement between the measurements and the model as well.

\section*{Off-diagonal long-range order}

BEC is associated with spontaneous breaking of the U(1) symmetry via emergence of a well-defined global phase. In our case, this would correspond to a buildup of spatial coherence. The size, dispersion and dimension of the system affect BEC in profound ways. First, the dispersion and dimension determine the occupation probabilities of excited states and thereby the possibility of BEC~\cite{Griffin1995,ketterle_bose-einstein_1996}. Second, fluctuations allow only quasi-long-range order in dimensions smaller than three~\cite{Kosterlitz2016}. It is thus essential to inspect the dimensionality, size, and dispersion of our system. SLR modes are known to be tightly confined in the third dimension, but whether our condensate is 1D or 2D is not determined a priori. By direct 2D $k$-space imaging we have confirmed that the condensate is confined in two dimensions, see Fig.~\ref{figure3}e. Above we have shown in detail how the system thermalizes when the population is propagating along the $y$-axis (polarization along $x$), following the SLR dispersion in $k_y$. The dispersion in the $x$-direction is parabolic, and the dispersion has a 2D minimum at ${\bf k}=0$ ($\Gamma$-point) (Extended Data Fig.~9). The dispersions (an example shown in Fig.~\ref{figure1}c) at high energies are linear, but around the band bottom, where the condensation occurs, they are flat and can be approximated for instance by a parabola. For 1D and 2D parabolic and 1D linear dispersions, a BEC is possible only in a finite system. Our system is finite: the effective size is given by the SLR mode coherence length (45 $\mathrm{\mu m}$). Although true long-range order is a subtle issue in low dimensions~\cite{Kosterlitz2016}, especially in non-equilibrium systems~\cite{Altman2015}, off-diagonal long-range order over a finite area may emerge at the critical density.

\begin{figure}[h!]
  \centering
    \includegraphics[width=0.9\textwidth]{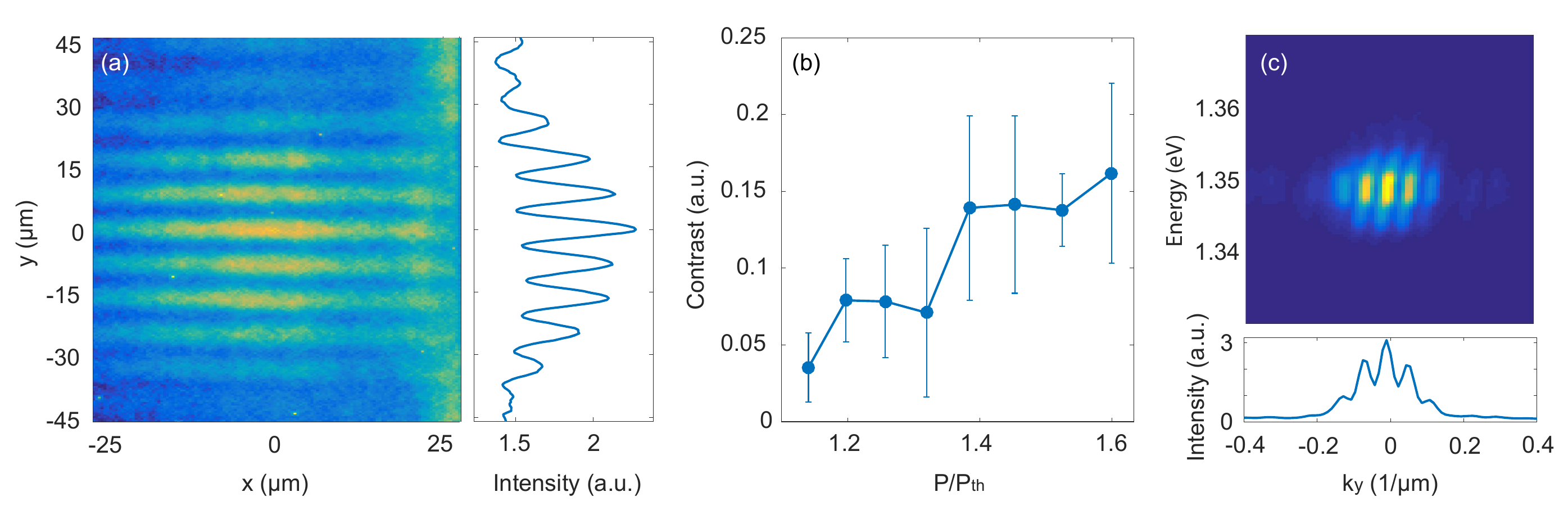}
   \caption{\footnotesize{
\textbf{Experiments on spatial coherence.} a) Retro-reflector image and the intensity profile (integrated over the $x$-direction) show that the condensate is spatially coherent over distances of at least 90 $\mathrm{\mu m}$. b) The degree of spatial coherence (contrast or visibility of the interference peaks) grows with increasing pump fluence $P$. The pump fluencies in the figure are above the threshold $P_{th}$. The corresponding results for the lasing experiment are presented in Extended Data Fig.~10. As an additional independent test of long-range order, we measured the emission from the sample through a double slit: c) shows the intensity measured through a double slit targeted on the sample position where the condensation occurs. Each slit had a width of 30 $\mathrm{\mu}$m and the inter-slit distance was 90 $\mathrm{\mu}$m. The observed intensity pattern is typical for double-slit interference showing that the emission source in the array must be coherent at least up to the inter-slit distance of 90 $\mathrm{\mu}$m, confirming the result given by the Michelson interferometer experiment. }}
   \label{figure4}
\end{figure}

We measured the spatial coherence by a Michelson interferometer setup with one of the mirrors replaced by a retro-reflector that inverts the image along the y-axis, see Methods. Combining the real-space image of the sample with its mirror image reveals the phase correlations between different parts of the sample. From the measured pattern shown in Fig.~\ref{figure4}a, we observe that the condensate is spatially coherent over distances of at least 90 $\mathrm{\mu}$m, confirmed by Fig.~\ref{figure4}c. This is twice as long as the natural coherence length of the SLR modes (45  $\mathrm{\mu}$m), demonstrating that the degree of spatial coherence has increased.
With increasing pump fluence above the threshold value, we see a gradual build-up of the degree of spatial coherence as shown in Fig.~\ref{figure4}b. 

\section*{Conclusions}

We studied nanoparticle arrays with dye molecules acting as a thermal bath, and monitored propagation of the surface lattice resonance (SLR) excitations after pumping. Fit to the BE distribution at room temperature, spectral narrowing, non-linear behaviour at critical density, as well as the onset and growth of spatial coherence were observed in our finite-size, open-dissipative system. Difference to lasing was demonstrated by suppressing the possibility of the population reaching the band edge via thermalization. The evolution of the band edge peak and thermalizing population revealed that lasing occurs at the pump spot while the BE distribution emerges from thermalizing population. These observations constitute evidence for a BEC of surface plasmon polariton lattice modes.

The sub-picosecond dynamics in our system is faster than the typical condensation dynamics in magnon condensates~\cite{demokritov_boseeinstein_2006} (100 ns), photon condensates\cite{klaers_bose-einstein_2010,schmitt_kinetics_2015,marelic_spatiotemporal_2016} (10 ps -- 1 ns) and most inorganic polariton condensates~\cite{deng_exciton-polariton_2010,carusotto_quantum_2013,byrnes_exciton-polariton_2014} (1--100 ps). It is  comparable to those in the fastest inorganic and organic polariton systems~\cite{baumberg_2008,daskalakis_nonlinear_2014,plumhof_room-temperature_2014} where sub-picosecond decay and coherence times have been observed. Our experiment allows monitoring the actual evolution of the thermalization and condensate formation in detail, as exemplified by Fig.~\ref{figure2}, something that has earlier only been possible in longer time-scales. 

Strong coupling has been achieved in plasmonic systems at room temperature~\cite{torma_strong_2015,chikkaraddy_single-molecule_2016}, and the SLR modes can hybridize with the molecules~\cite{vakevainen_plasmonic_2014}. This may provide effective interactions and thus a new arena for photonic quantum fluids. The lattice geometry can be changed into more complex ones, such as hexagonal~\cite{guo_geometrydependence_2017}, in search for degeneracy (Dirac) points. Customizable lattice geometries are exciting, especially because recent discoveries of topological phenomena~\cite{hasan_colloquium_2010} have accentuated the significance of the band structure. Symmetries can be broken, for instance, by nanoparticle shape or use of magnetic material~\cite{kataja_surface_2015} to explore how condensation may be transformed in a topologically non-trivial system. Finally the new type of BEC we have created is highly accessible technologically: it is based on on-chip systems that can be scaled up by using nanolithography and microfluidistics, and operate at room temperature.\\

\noindent \textbf{Online Content} 
Methods, along with any additional Extended Data display items and 
Source Data, are available in the online version of the paper; references unique to 
these sections appear only in the online paper.

\section*{Methods}
\subsection*{Sample fabrication}

Square arrays of rod shaped gold nanoparticles were fabricated on a borosilicate substrate by electron-beam lithography and subsequent electron-beam evaporation of 2 nm of titanium and 50 nm of gold. The array periodicities ranged from $p_x = p_y =$ 580 nm to 610 nm. The width of the particles was fixed to 100 nm and the length was chosen to be 65 $\%$ of the period $p_y$ to provide a sufficient band gap in the dispersion. The IR-792 molecules were dissolved into a solution of 1:2 (dimethyl sulfoxide):(benzyl alcohol). The time-resolved photoluminescence (TRPL) of the molecule does not show any wavelength dependence (Extended Data Fig.~11) that would complicate the analysis of the results.

\subsection*{Measurement setup}

The $x$-axis of the 2D charge coupled device (CCD) camera of the spectrometer was used to resolve the wavelength. Depending on the measurement scheme, the $y$-axis was used to resolve either 1) the angle of the transmitted/emitted light or 2) the $y$-position along the array. The spectra were collected over 300 ms integration time, that is, averaged over 300 laser pulses.

\subsubsection*{Angle-resolved measurements}
Angle-resolved transmission measurements (as in Fig.~\ref{figure1}c) were obtained by focusing light from a halogen lamp onto the sample and collecting the transmitted light with an objective (10X, 0.3 NA). The back focal plane of the objective was focused to the entrance slit of the spectrometer, and thus each position along the $y$-axis of the CCD corresponded to a particular angle of transmitted light. The $E(k)$ dispersions were subsequently calculated from the angle and wavelength-resolved spectra as $E=hc/\lambda_0$ and $k=k_0 sin(\theta)$. Here $k_0=2\pi/\lambda_0$ and $\theta$ is the angle with respect to sample normal. The dispersions show a band edge, and we have confirmed that the band-edge energy remains the same (with accuracy of four digits in eV) throughout the array (Extended Data Fig.~12).

Spatially specific k-space images of the sample emission (such as in Fig.~\ref{figure3}), were measured in a similar fashion, but the sample was pumped with a pulsed femtosecond laser (100 fs pulse duration, 750 nm center wavelength, 1 kHz repetition rate, $\mathrm{\mu}$J pulse energy), with a flat pulse profile. Pulsed excitation allows for high photon densities necessary for quantum degeneracy, while simultaneously avoiding bleaching of the organic molecules. Further, an iris was placed on the image plane of the sample to spatially filter only the light emerging from a specific region of the sample. For the spatial coherence measurements, the iris was replaced with a double slit having a 90 $\mathrm{\mu}$m inter-slit distance and 30 $\mathrm{\mu}$m slit width. For a schematic, see Extended Data Fig.~13a.

\subsubsection*{Spatially resolved measurements}
For spatially resolved spectra such as in Fig.~\ref{figure2}, the real space image of the sample was focused to the spectrometer entrance slit. Thus the $y$-axis of CCD corresponded to a particular position along the $y$-axis of the array while the $x$-axis corresponded to the wavelength. The magnification of the real space image was 4x at the entrance slit and the entrance slit width was 300 $\mathrm{\mu m}$ so that we collected the luminescence from 3/4 of the 100 $\mathrm{\mu m}$ array (in $x$-direction). For a schematic, see Extended Data Fig.~13b.

\subsubsection*{Michelson interferometer}
Spatial coherence properties were studied with a modified Michelson interferometer configuration, for a schematic see Extended Data Fig.~13c. The luminescence of the sample was first passed through a 900 nm (1.378 eV) long pass filter and then was directed to a non-polarizing beam splitter. One output arm of the beam splitter consisted of a hollow roof L-shaped mirror used to invert y to –y upon reflection. The other arm consisted of a regular mirror. The two reflected beams were directed through the beam splitter onto a lens and then onto a CCD. The first-order correlation function describing the degree of spatial coherence reads $g^{(1)}(\boldsymbol{y}, \boldsymbol{y}') = \frac{\langle E^*(\boldsymbol{y}) E(\boldsymbol{y}') \rangle }{\langle E^*(\boldsymbol{y}) \rangle \langle \langle E(\boldsymbol{y}') \rangle}$ (here $E^*(\boldsymbol{y})$ is the electric field at point $\boldsymbol{y}$). The contrast is obtained directly from the interference pattern as 
\begin{align}
\label{contrast}
C(\boldsymbol{y}, -\boldsymbol{y}) &= \frac{I_\textrm{max}-I_\textrm{min}}{I_\textrm{max}+I_\textrm{min}} = \frac{2 \sqrt[]{I(\boldsymbol{y})I(-\boldsymbol{y})}}{I(\boldsymbol{y})+I(-\boldsymbol{y})} g^{(1)}(\boldsymbol{y}, -\boldsymbol{y}).
\end{align}
Here, $I(\boldsymbol{y})$ is the light intensity and $I_\mathrm{max}$ and $I_\mathrm{min}$ are the maximum and minimum intensities of the fringes, respectively. Note that in our case the contrast is expected to decrease for increasing $y$ due to the decay ($I(\boldsymbol{y})<I(-\boldsymbol{y})$) of the SLR population, therefore the contrast gives only a lower limit to the value of $g^{(1)}(\boldsymbol{y}, -\boldsymbol{y})$.

\subsection*{Thermalization rate measurement}
The dependence of thermalization rate on concentration (Fig.~\ref{figure2}e) was determined by repeating the thermalization experiment of Fig.~\ref{figure2}a with different concentrations (25, 50, 75, 100, and 150 mMol). From the intensity spectra (as Fig.~\ref{figure2}a-c), we then obtained the slope of the red shift by fitting a line to the maxima of the intensity at each $y$ position. The slope of the red shift interprets as the rate of the thermalization process by scaling the distance with the SLR-mode group velocity. The slopes were measured from the energy regime that is far from the band edge, i.e. from the linear part of dispersion. Various periodicities were used in the experiments. The data in Fig.~\ref{figure2}e shows the mean value of several (3--7) measurements for each concentration, with error bars showing the standard deviation within each data set. Note that the linear fits for thermalization rate are valid in a limited energy range. This is evident from Extended Data Fig.~14 where the thermalization and emergence of BEC is shown for a period $p =  600$ nm sample with a broader energy range (at the expense of spectral resolution).

\subsection*{Theoretical model}
We used a rate-equation model~\cite{Kirt13} that describes population dynamics but neglects correlations relevant for strong coupling. In the model, an ensemble of spectrally broadened two-level dye molecules is pumped by a laser pulse. The dye molecules interact with discretized, lossy SLR modes by emission and re-absorption. Measured emission and absorption profile taken from literature, as well as realistic parameters for the loss rates and lifetimes were used in the model. Parameters in Eq. 2 of Supplemental Information, used in the simulations are the following: SLR decay rate $\kappa = 1.08 \times 10^{13}$ 1/s, number of molecules is varied $N = 1.18...7.08 \times 10^7$ ($N = 2.34 \times 10^7$ corresponds to experiments with concentration of 50 mMol), coupling constant $g = 10^{4.5}$, pump amplitude is varied $P = 5.01 \times 10^{10} ...13.80 \times 10^{10}$, and spontaneous decay rate for the molecules $\gamma_m = 10^9$ 1/s. With these parameters, the fraction of molecules that is excited by the pump is small, therefore there is a large number of molecules in thermal equilibrium available for re-absorption processes.
The main approximations compared to a full microscopical description are that a phenomenological parameter (the coupling $g$) is used for the overall strength of absorption and emission, and the field profiles are assumed uniform.
There are no computational states in the region of the energy band gap of the dispersion, which produces an empty area in Fig.~\ref{figure2}d and Fig.~\ref{figure3}d. The calculation of the thermalization time as a function of molecular concentration, Fig.~\ref{figure1}e, constitutes a self-consistency check for the correspondence between the experiment and the model. Namely, we \textit{varied the concentration while keeping all other parameters fixed}, and obtained the good agreement between the model and the measurements shown in Fig.~\ref{figure2}e. For detailed formulation of the model, see Supplementary Information. 

\subsection*{Fits to Bose--Einstein and Maxwell--Boltzmann distributions}

For the measured population distributions, examples shown in Fig.~\ref{figure3} and Extended Data Fig.~6, fits to Bose--Einstein (BE) and to Maxwell--Boltzmann (MB) distributions were made using non-linear least squares fitting. In a few meV vicinity of the $\Gamma$-point the dispersion is quite flat both in $x$- and $y$-directions, and for higher energies it is essentially linear but anisotropic in the two dimensions as shown in Extended Data Fig.~9. The density of states used in fitting was obtained numerically from such dispersion. Above threshold, excellent fits to the BE distribution around room temperature were obtained while the MB distribution did not fit well. Below threshold, the MB distribution gives good results at $T=300$ K. To obtain statistics for the temperature of the condensate, we made the BE distribution fit to data obtained by averaging over ten separate measurements. All the measured population distributions were taken using the corresponding pump fluencies and from the same region of the array (the red circle in Fig.~\ref{figure3}). 
The resulting fit to the BE distribution gives temperature of $T=269 \pm 67$ K, error limits showing 95\% confidence bounds for the fit. With 99\% confidence bounds the result is $T=269 \pm 89$ K. In contrast, for the sample exhibiting lasing the fit to the Bose--Einstein distribution gives a temperature of only 25 K, see Extended Data Fig.~2, reflecting the non-equilibrium character of lasing and lack of thermalization to the environment.
\\

\noindent \textbf{Acknowledgements} We thank Miikka Heikkinen, Dong-Hee Kim, Robert Moerland and Marek Ne\v cada for useful discussions. This work is dedicated in memory of D. Jin and her inspiring example. This work was supported by the Academy of Finland through its Centres of Excellence Programme (2012–-2017) and under project NOs. 284621, 303351 and 307419, and by the European Research Council (ERC-2013-AdG-340748-CODE). This article is based upon work from COST Action MP1403 Nanoscale Quantum Optics, supported by COST
(European Cooperation in Science and Technology). K.S.D. acknowledges financial support by a Marie Skłodowska-Curie Action (H2020-MSCA-IF-2016, project id 745115). Part of the research was performed
at the Micronova Nanofabrication Centre, supported by Aalto University. Triton cluster at Aalto University was used for the computations.  
\\

\noindent \textbf{Author contributions} P.T. initiated and supervised the project. T.K.H., A.J.M., R.G., and A.I.V. performed the experiments. A.J.M., T.K.H., and A.I.V. analysed the data. T.K.H., K.S.D., and H.T.R. built the experimental setup. A.J.M., J.-P.M., and A.J. performed the theoretical modeling. R.G. fabricated the samples. All authors discussed the results. P.T., A.J.M., and T.K.H. wrote the manuscript together with all authors.
\\

\noindent \textbf{Additional information}
Supplementary information is available in the online version of the paper. Reprints and
permissions information is available online at www.nature.com/reprints. Correspondence and requests for materials should be addressed to P.T. \\

\noindent \textbf{Competing financial interests}
The authors declare no competing financial interests.\\

\bibliography{bibliography_thermalization_no_doi}

\newpage

\noindent \textbf{Extended Data Figures and Captions}\\

\begin{figure}[h!]
  \centering
    \includegraphics[width=0.72\textwidth]{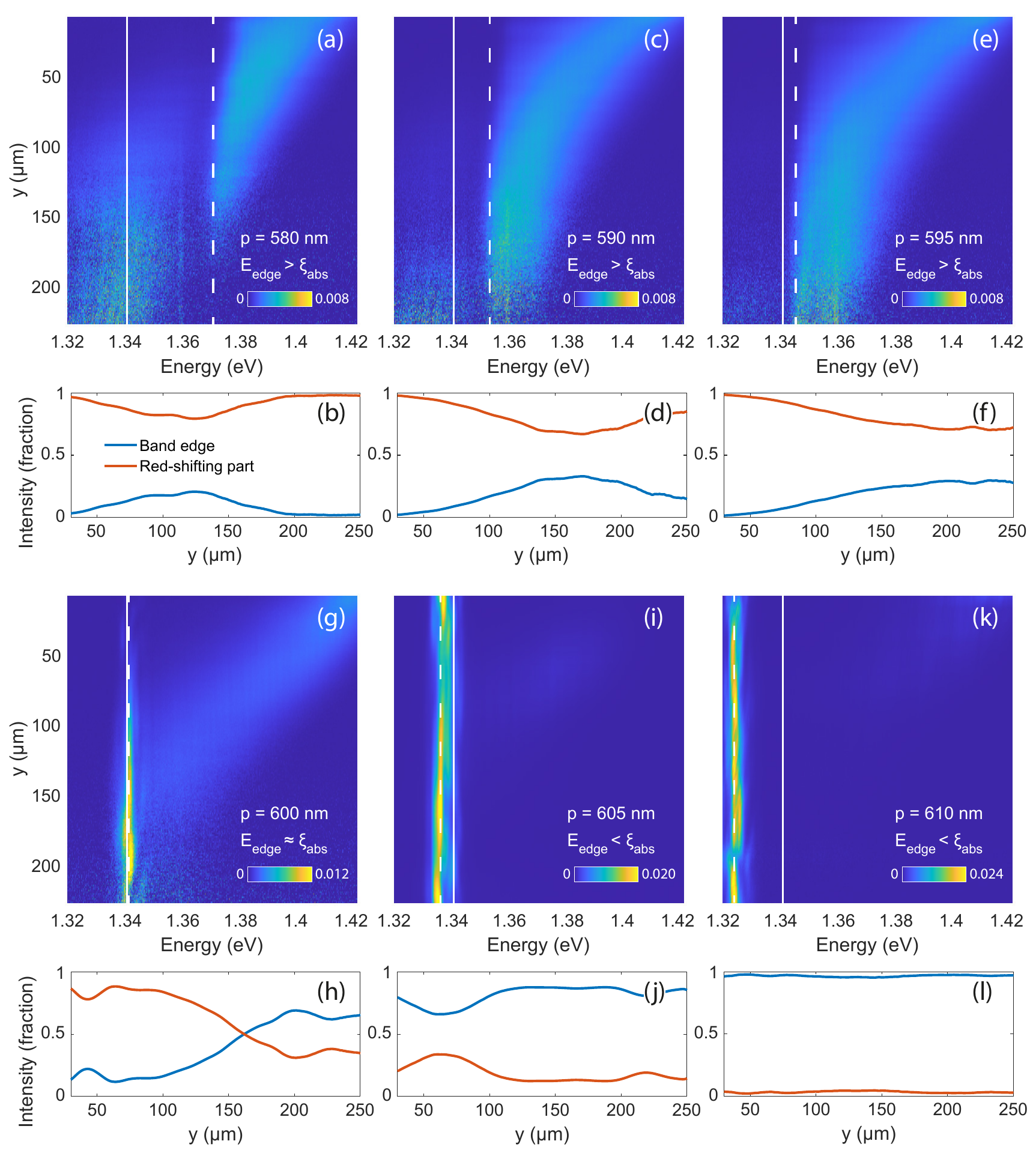}
   \caption*{{\bf Extended Data Figure 1.} Spectral and spatial evolution of the sample luminescence. The change from thermalization to BEC to lasing is realized by varying the periodicity $p$ of the sample: $p =$580 nm (a, b), 590 nm (c, d), 595 nm (e, f), 600 nm (g, h), 605 nm (i, j) and 610 nm (k, l). Here $\xi_{\mathrm{abs}}$ and the band edge energy are marked with a solid and dashed lines in (a), (c), (e), (g), (i), (k), respectively. For $p =$580 nm, the band edge population gradually increases upon propagation, and due to significant reabsorption and -emission, leaks below the band edge. The gradual buildup and decay of the band edge population is evident from the blue curve in (b). For $p =$590 nm, the band edge cumulates a higher population, which again leaks below the band edge. For 595 nm, the band edge almost overlaps with $\xi_{\mathrm{abs}}$. Due to reduced absorption at the band edge energy (and therefore suppressed re-emission), there is no significant red shift of the population once the band edge has been reached. This is also evident in (f), where the band edge population first increases, then stays constant. For 600 nm, the band edge equals $\xi_{\mathrm{abs}}$. The red-shifting population accumulates to the band edge, and due to negligible absorption and re-emission probability, a macroscopic occupation (70\%, see (h)) and BEC is formed. With 605 nm, there is a significant population at the band edge already in the beginning of the array due to stimulated emission (lasing). Additionally, a very faint signal from red-shifting population can be seen at $y=$70$\mu$m, $E =$1.38 eV). This can be considered as intermediate case between BEC and lasing: while stimulated emission governs the system dynamics, the red-shifting population is still non-negligible (10-20 \%, see (j)). Finally, for 610 nm, typical lasing behaviour is seen: 1) the relative intensity of the red-shifting part is essentially zero, implying that absorption and re-emission processes do not contribute to the dynamics, and 2) macroscopic population is visible at the band edge energy throughout the array.
Colour scales in the false colour plots are saturated such that the maximum intensity values are (a) 0.0096, (c) 0.0130 (e) 0.0112, (g) 0.0175, (i) 0.0231 and (k) 0.0286.}
\end{figure}

\newpage

\begin{figure}[h!]
  \centering
    \includegraphics[width=0.8\textwidth]{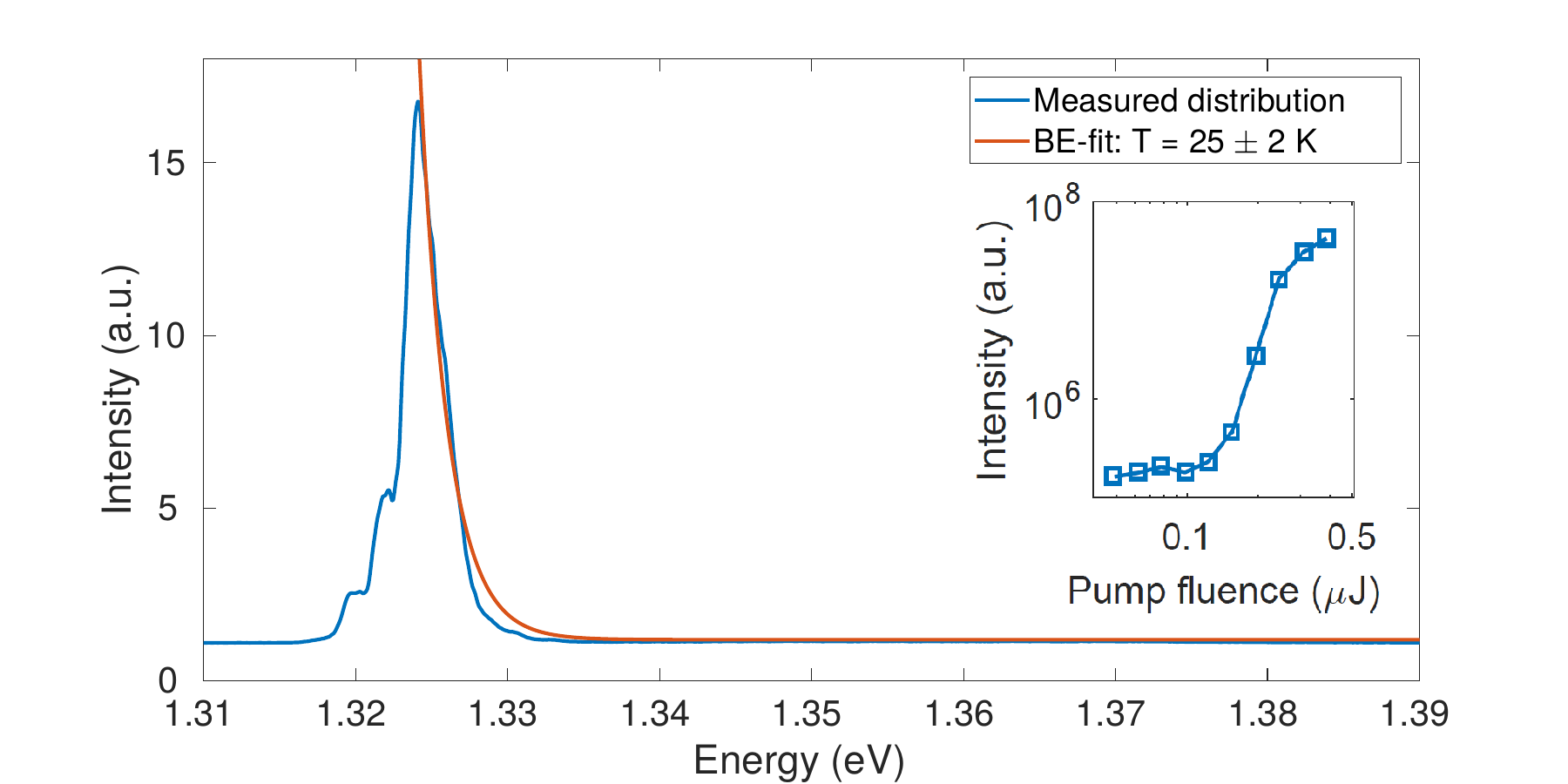}
   \caption*{{\bf Extended Data Figure~2.} Measured intensity distribution for the 610 nm period sample that shows lasing. Strongly nonlinear luminescence intensity is observed with increasing pump fluence, see the inset. The thermalized tail is vanishingly small in the observed population distribution. Consequently, the Bose--Einstein (BE) distribution fit gives an unrealistically low temperature of 25 $\pm$ 2 K (95\% confidence limits; the 99 \% confidence limits are $\pm$ 3 K).} 
\end{figure}

\newpage

\begin{figure}[h!]
  \centering
    \includegraphics[width=0.6\textwidth]{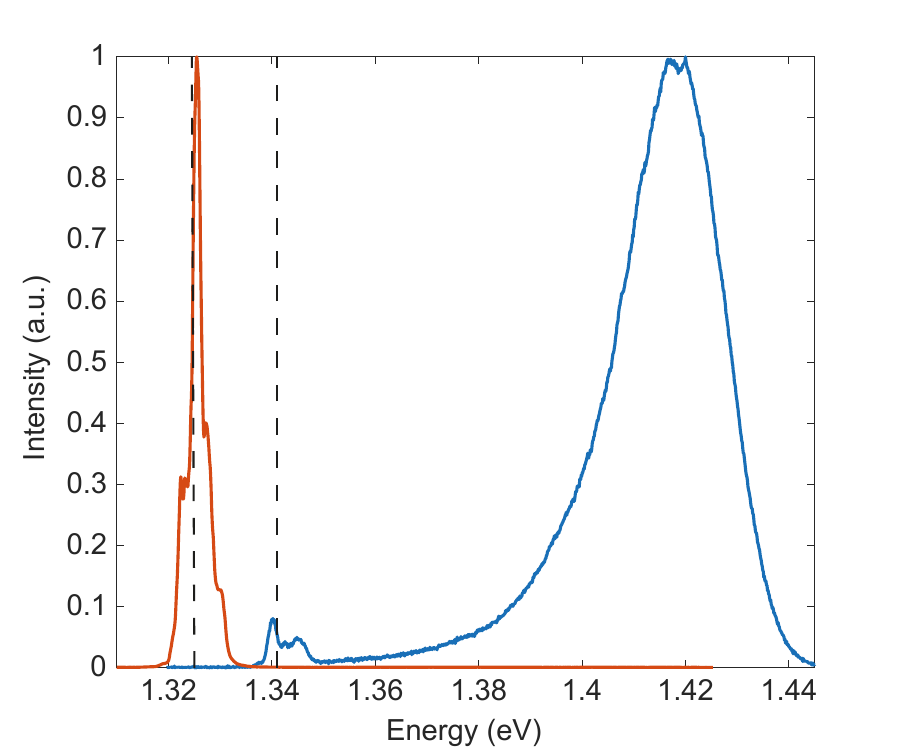}
   \caption*{{\bf Extended Data Figure 3.} The emission spectra collected from the part of the array that is under the pump spot, for the BEC (blue curve) and lasing (red curve) case. For the BEC case the emission mainly takes place at energy that corresponds to the emission maximum of the dye molecule. A very small emission peak is observed at the band-edge energy (dashed line). In contrast, for the lasing case, the emission maximum is observed at the band-edge energy (dashed line).}
\end{figure}

\newpage

\begin{figure}[h!]
  \centering
    \includegraphics[width=0.9\textwidth]{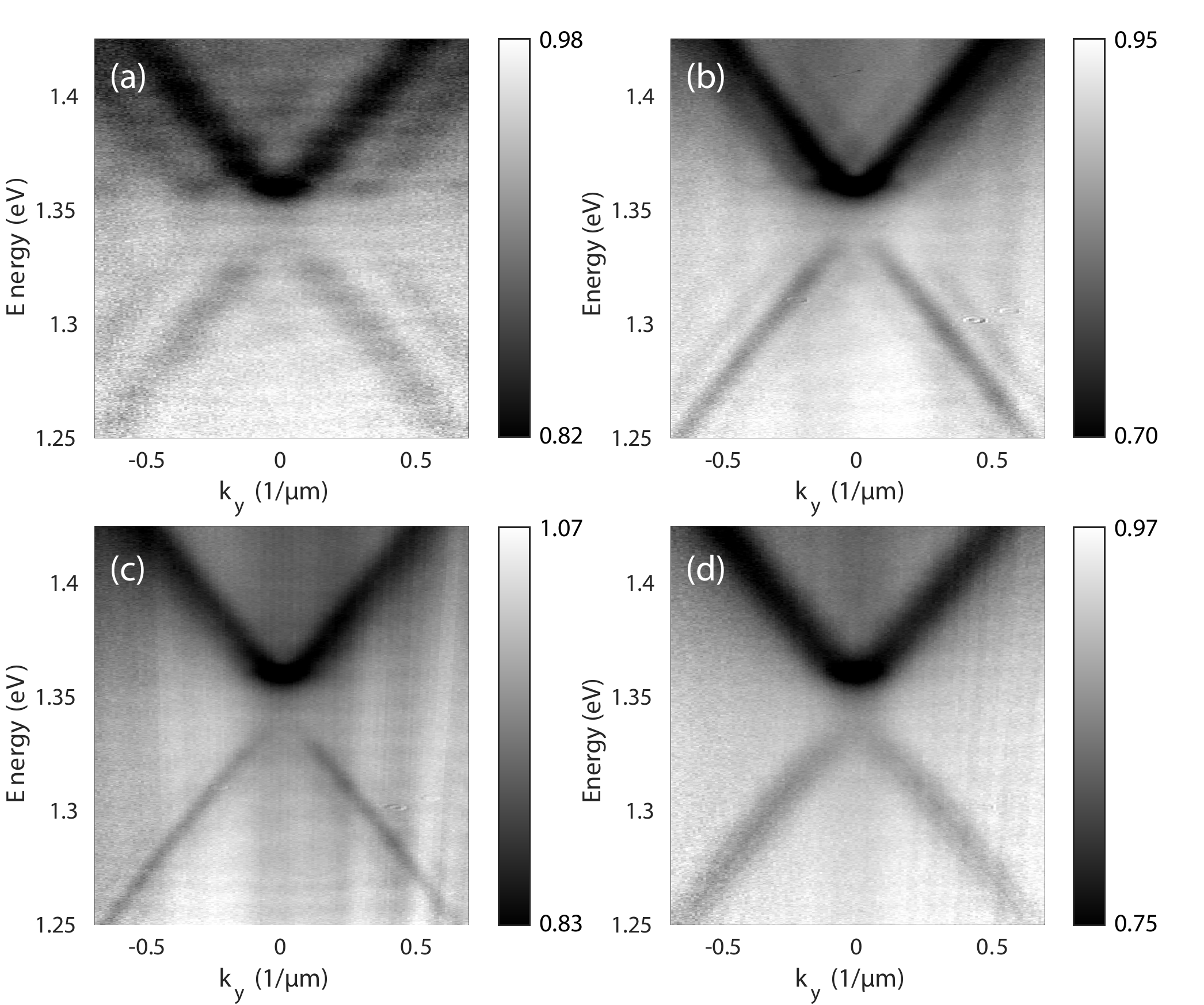}
   \caption*{{\bf Extended Data Figure 4.} Double slit experiments for a nanoparticle array with periodicity corresponding to the BEC case. These measurements were carried out in the absence of molecules by transmitting white light through the sample and then through a double slit. The used inter-slit distances were 15, 30, 45 and 90 $\mathrm{\mu} \mathrm{m}$ (a, b, c and d, respectively). A clear beating pattern indicating spatial coherence is seen with the smaller inter-slit distances. However, in (d) no interference is visible, in contrast to the clear interference fringes observed for the same inter-slit distance in the presence of molecules (Figure 4c of the main text), highlighting the increased spatial coherence in the BEC. Gray scales are saturated for both high- and low-end values to find the optimal visibility of interference fringes. The minimum and maximum transmission values are as follows (a) $\rm T_{min} = 0.77$, $\rm T_{max} = 1.08$; (b) $\rm T_{min} = 0.58$, $\rm T_{max} = 1.05$; (c) $\rm T_{min} = 0.71$, $\rm T_{max} = 1.14$; and (d) $\rm T_{min} = 0.67$, $\rm T_{max} = 1.06$. The values above unity are due to noise and imperfect normalization with respect to the incident light spectrum.}
\end{figure}

\newpage

\begin{figure}[h!]
  \centering
  \includegraphics[width=0.65\textwidth]{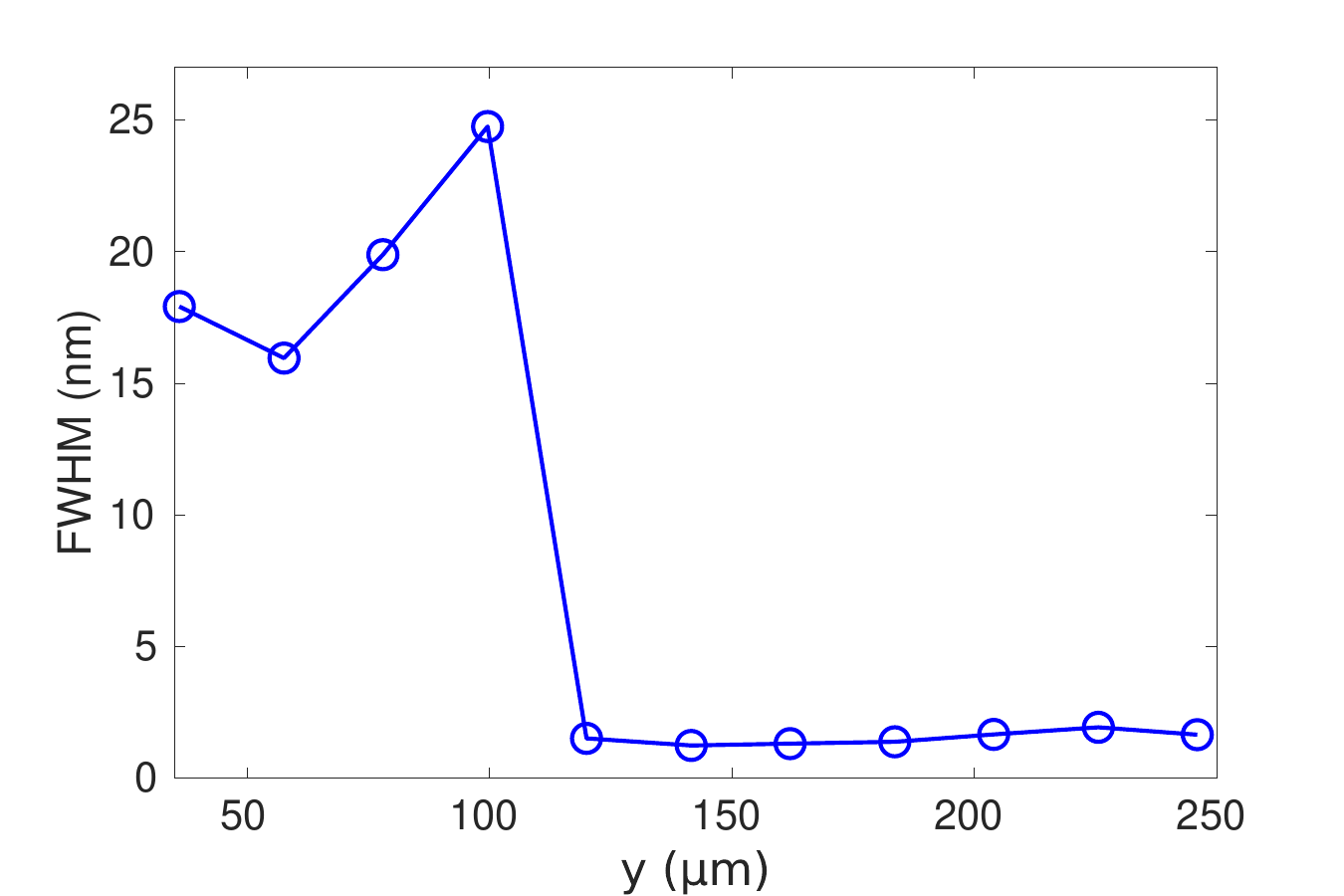} 
   \caption*{{\bf Extended Data Figure 5.} Full width at half maximum (FWHM) as a function of position along the array, obtained from the BEC case ($p = 600$ nm) presented in Fig.~2c. The FWHM is measured from the peak that has the highest intensity at each position in y, that is, from the thermalizing cloud until a higher intensity feature appears at the band edge.}
\end{figure}

\newpage

\begin{figure}[h!]
  \centering
   \includegraphics[width=0.85\textwidth]{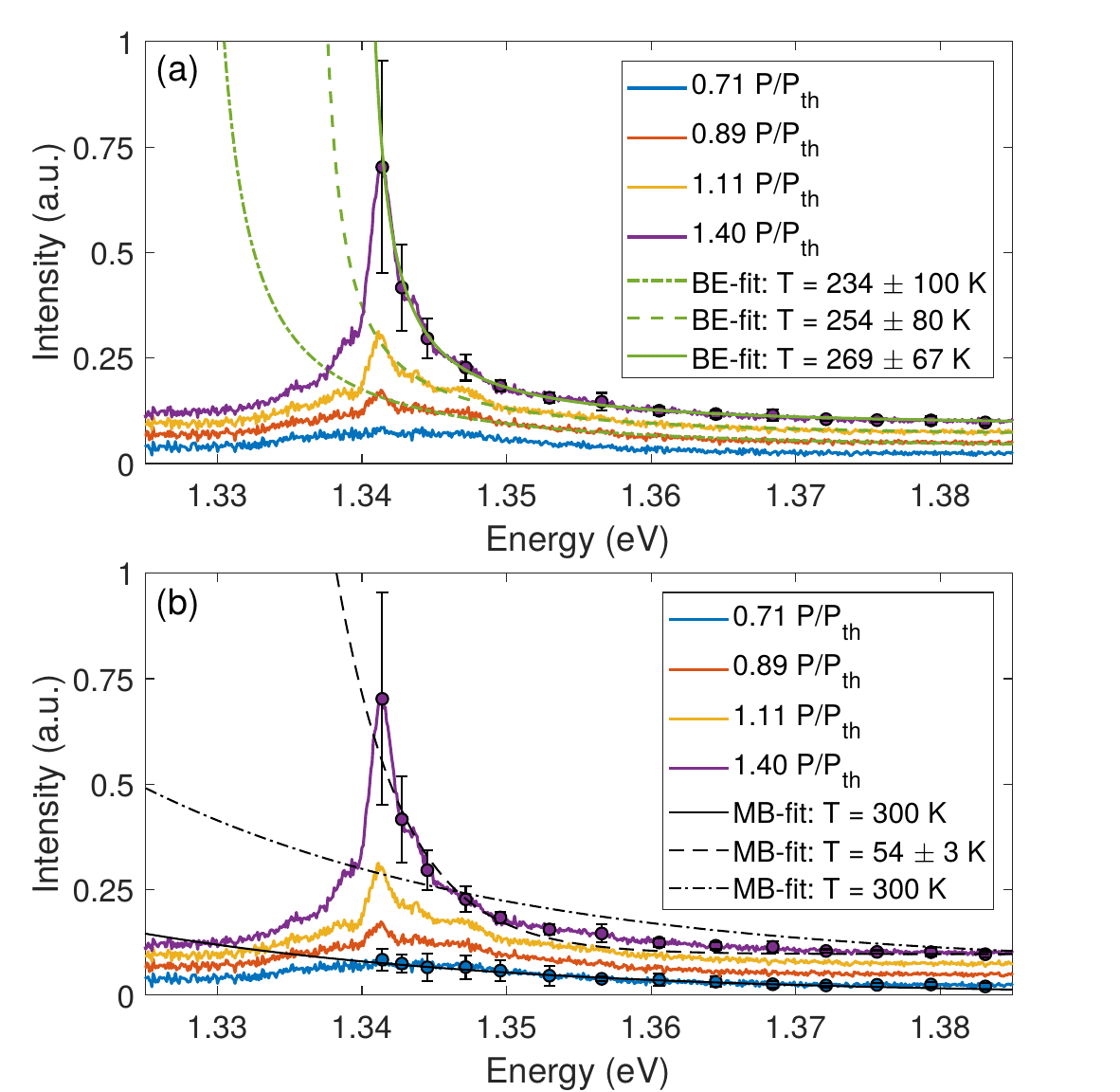}
   \caption*{{\bf Extended Data Figure 6.} (a): Measured intensity distributions with various pump fluencies for the 600 nm periodicity sample, where BEC is observed with the highest pump fluence (purple solid line). (a) Fits to the Bose--Einstein distribution for lower pump fluencies. The fits show that the BE distribution emerges at the threshold, with peak at the band edge. The fit to the BE distribution at second-highest pump fluence (dashed line) gives temperature of 254 $\pm$ 80 K (95 \% confidence level; 99 \% confidence bounds are $\pm$ 105 K). The fit to the third-highest pump fluence (dash-dot line) gives temperature of of 234 $\pm$ 100 K (95 \% confidence level; 99 \% confidence bounds are $\pm$ 132 K), but as the peak diverges far from the band-edge, it is evident that the critical photon density has not been reached. For the lowest pump fluence, 0.71$P_{th}$, the signal-to-noise is too low for a proper fit to the BE distribution. (b) Fits to the Maxwell--Boltzmann (MB) distribution for measured distributions. The MB distribution with temperature fixed to 300 K
(dash-dot line) fits very poorly with the measured distribution at the highest pump fluence. By including the temperature as a fitting parameter, the agreement is somewhat better, but the temperature of 54 $\pm$ 3 K (95 \% confidence level; 99 \% confidence bounds are are $\pm$ 4 K) (dashed line) obtained from the fitting is unrealistically low. Expectedly, at low pump fluencies below BEC threshold the experimentally measured distribution (blue line) fits better with the MB distribution with temperature fixed to 300 K (solid line).}
\end{figure}  

\newpage

\begin{figure}[h!]
  \centering
    \includegraphics[width=0.82\textwidth]{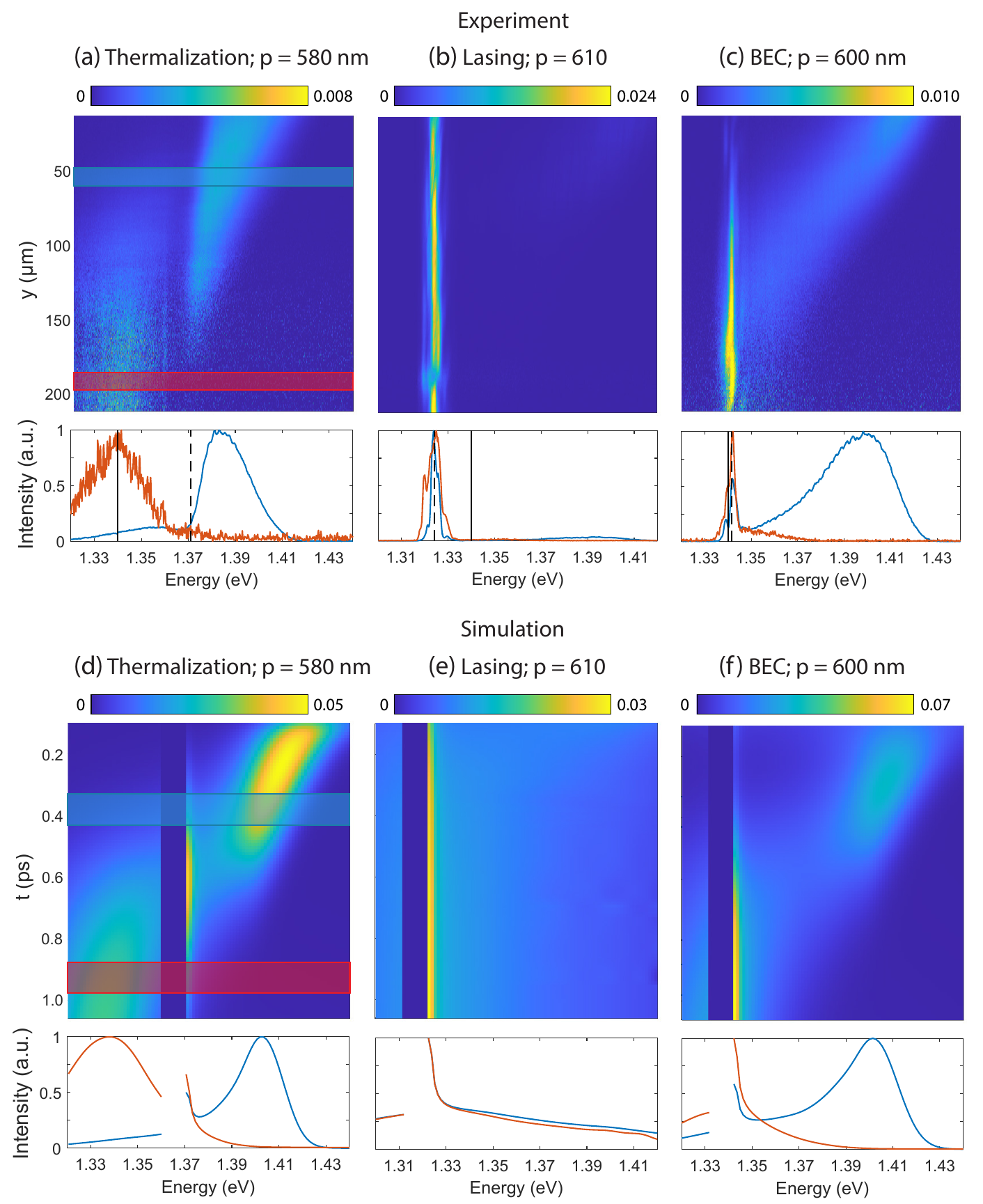}
   \caption*{{\bf Extended Data Figure 7.} Experimental spectral evolution of the sample luminescence and the corresponding computational results for different periodicities. The first, second and third columns correspond to samples having periodicities 580, 610 and 600 nm, respectively. The first row (a-c) shows the spatially resolved experimental spectra, which are normalized with the integrated spectral intensity along each row. The second row (d-f) shows the corresponding simulation data. The lower panels in (a-c) show the normalized intensity distributions obtained from different $y$-positions along the sample. The blue curves are obtained by summing the intensities over the range $y$ = 48--60 $\mathrm{\mu m}$ (indicated by the blue bounding box in (a)) and the red curves are obtained by summing over $y$ = 186--198 $\mathrm{\mu m}$ (the red bounding box in (a)). The corresponding distributions from simulations are shown in the lower panels of (d-f) and are obtained by summing the intensities over $t$ = 0.33--0.43 ps (blue curves) and $t$ = 0.88--0.98 ps (red curves). The ranges of summations are also marked by the blue and red bounding boxes in (d). The data in (a-c) is the same as in Fig.~2a-c and the saturation levels of the colour scale are the same. The simulation plots (d-f) do not require saturation of colour scales.
}
\end{figure} 

\newpage

\begin{figure}[h!]
  \centering
    \includegraphics[width=0.6\textwidth]{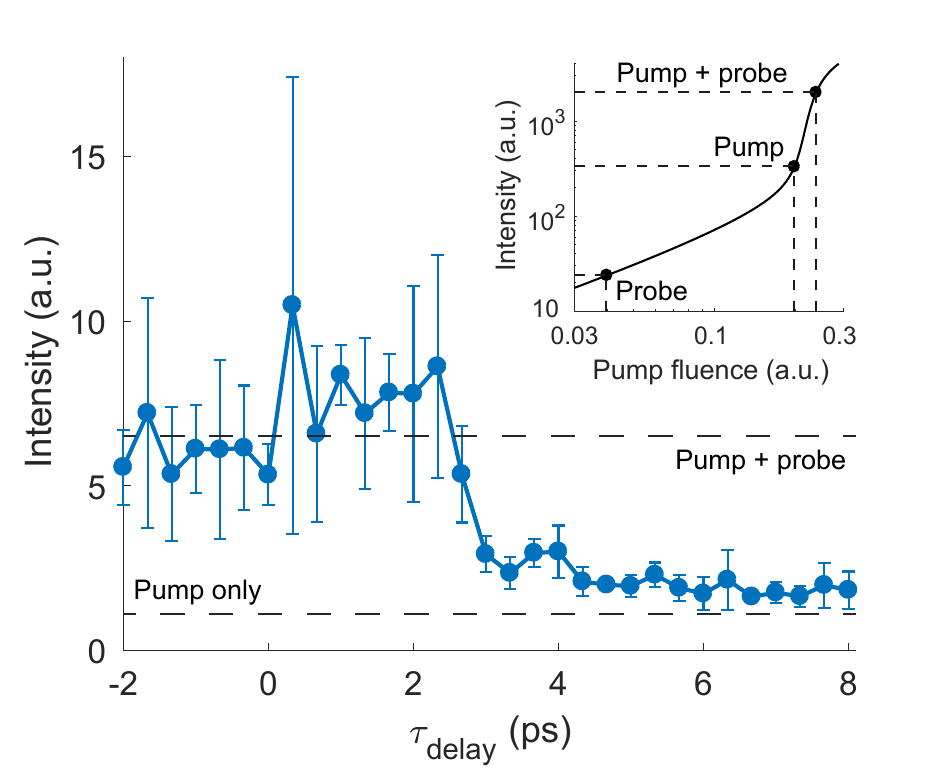}
   \caption*{{\bf Extended Data Figure 8.} Measured intensity as a function of pump-probe delay, confirming the ultrafast dynamics of the system. For the experimental setup of the pump-probe measurements see Extended Data Fig.~13d. The data points were obtained by taking the average of three measurements (the error bars represent the standard deviation). Positive delay times ($\tau > 0$) refer to the pump arriving at the sample before the probe. Similarly to all other experiments, the pump was focused to the edge of the array. The probe (diameter approximately 110 $\mathrm{\mu m}$) was focused to the center of the array (where the BEC occurs) and the luminescence intensity was measured from the same area. The pump intensity was chosen to be 1.1 times the threshold value for BEC and the probe intensity to be 20 percent of that of the pump. This ensures that 1) the pump induces a macroscopic population to the center of the array, and 2) the system is in the nonlinear-response regime, see the inset. In the nonlinear regime, the spatial and temporal overlap of the probe beam with the BEC is expected to result in nonlinear increase in the sample luminescence intensity.
At $\tau > 3$ ps, a constant but low intensity level is observed. This corresponds to the case where the pump-induced BEC has decayed before the probe arrives. Thus the molecules excited by the probe cannot contribute to the condensate population. (Due to long (300 ms) integration time of the spectrometer, the total measured intensity equals the sum of pump and probe induced luminescence intensities in the linear regime.)
With delays $\tau < 2.5$ ps, a significantly higher intensity level  is observed. This corresponds to the case where the weak probe signal excites the molecules at the center of the array prior to the arrival of the BEC pulse. Due to long (ns) spontaneous emission lifetime of the molecules, they stay in the excited state until the pump-induced BEC is formed. These molecules then emit photons to the BEC via stimulated emission. At delays 2.5 ps $< \tau <$ 3.0 ps  we observe a fast picosecond decay in the intensity, confirming the ultrafast dynamics of the BEC.}
\end{figure} 

\newpage

\begin{figure}[h!]
  \centering
    \includegraphics[width=0.7\textwidth]{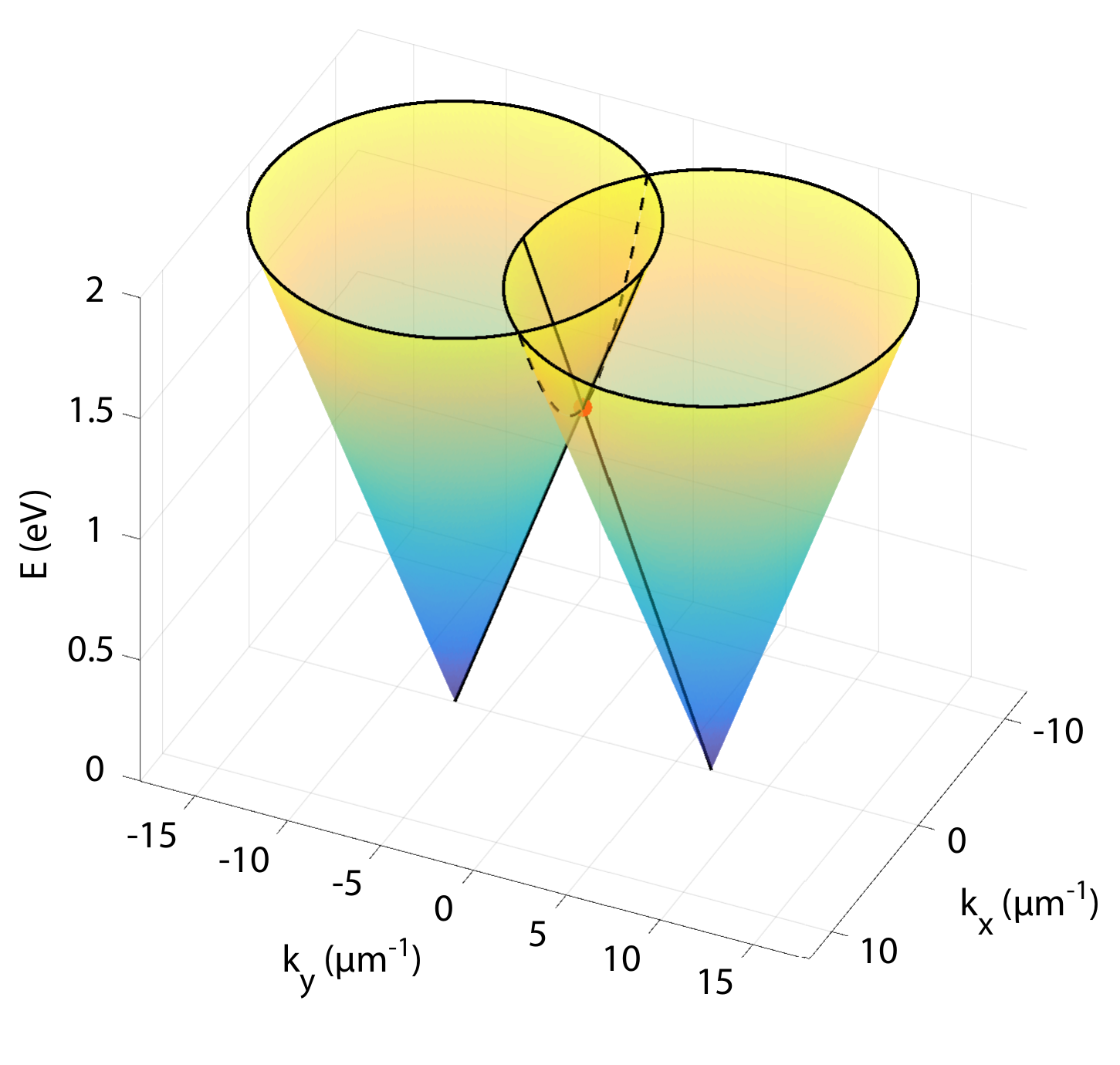}
   \caption*{{\bf Extended Data Figure 9.} Dispersion of the SLR mode closely follows the diffracted orders (DOs) of the lattice. Here, we have plotted the first DOs of light propagating in periodic structure in $y$-direction. In Fig.~\ref{figure1}c and thereafter, the dispersions are measured along $k_y$. The DOs intersect at the $\Gamma$-point ($\textbf{k}=0$) that is marked with a red dot. In $y$-direction the crosscut of the DOs is linear (solid lines) and in $x$-direction parabolic (dashed line). These two DOs combined with the x-polarized plasmonic resonances in individual nanoparticles result in the SLR mode in our structures. The band-gap opening at the $\Gamma$-point provides a two-dimensional band-bottom for the SLR excitations to condensate in.}
\end{figure} 

\newpage

\begin{figure}[h!]
  \centering
    \includegraphics[width=0.9\textwidth]{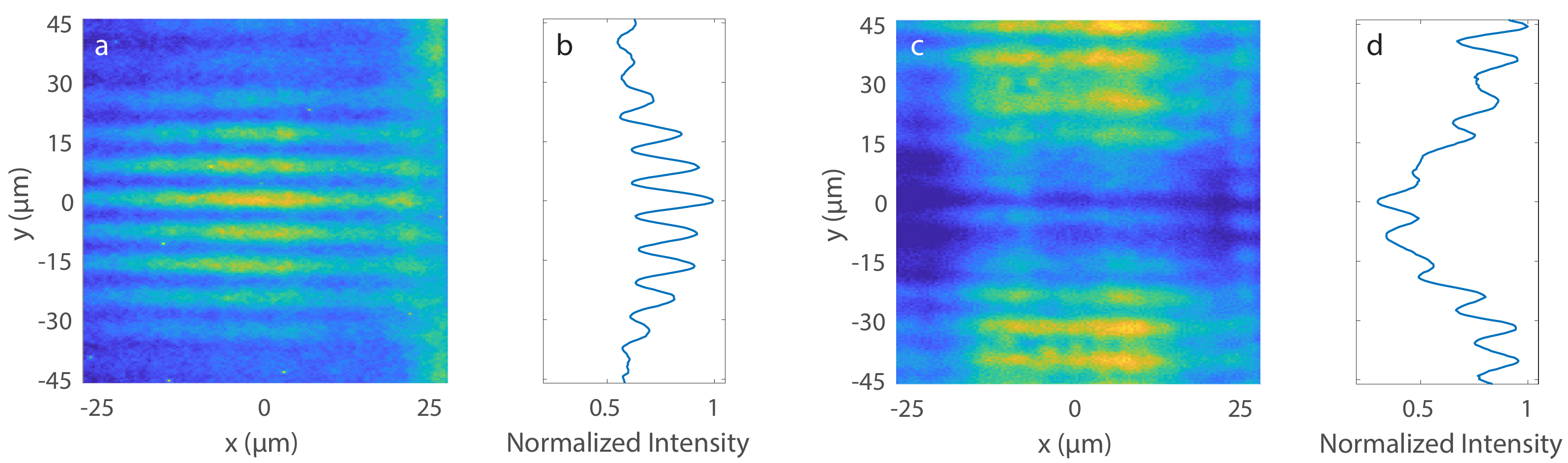}
   \caption*{{\bf Extended Data Figure 10.} Spatial coherence properties of BEC (a, b) and lasing (c, d). The different intensity distributions of the coherent populations in the two cases are evident from the different envelopes of the interference pattern: For BEC (a), the coherent population originates from the thermalization upon propagation: a retroreflector configuration is used to invert y to -y at the position where the coherent population maximizes. This produces a maximum of the envelope to the center of the image (b). For lasing, however, the intensity maximizes at the edge of the array, then decays with increasing distance from the edge. In this case the y to -y inversion (at the same position as for the BEC case) creates a minimum of the envelope to the center of the image (d).}
\end{figure}

\newpage

\begin{figure}[h!]
  \centering
  \includegraphics[width=0.8\textwidth]{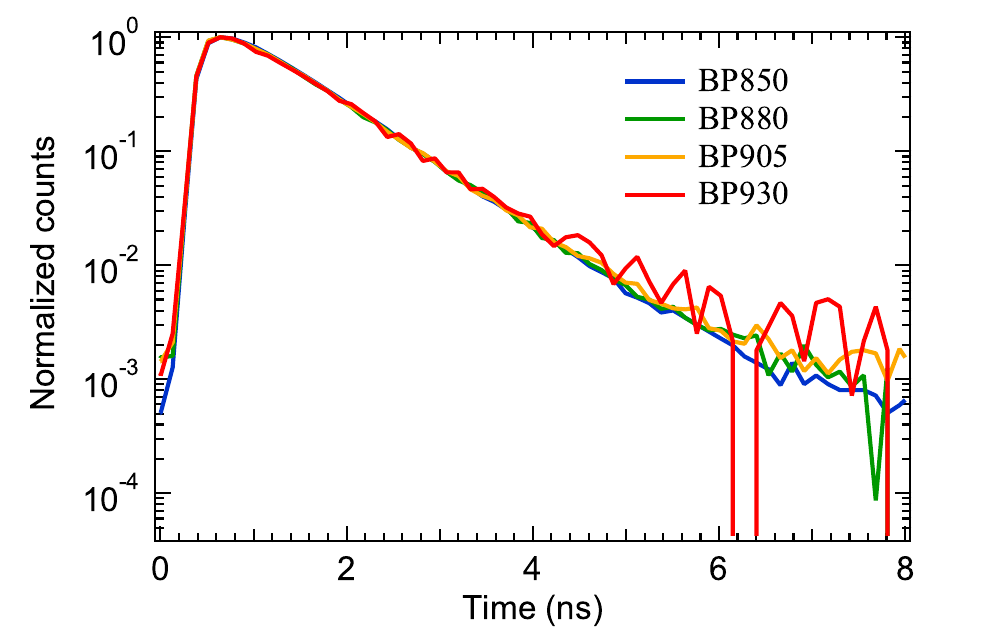} 
   \caption*{{\bf Extended Data Figure 11.} Time-resolved photoluminescence (TRPL) of a 2.5mM IR-792 solution, obtained by using a time-correlated single-photon counting (TCSPC) setup. The solution was excited with a 510~nm diode laser. Emission from the solution is spectrally resolved with bandpass filters and focused onto an avalanche photodetector (Micro Photon Devices) with 64~ps resolution. The measured decay histograms exhibit an exponential decay with lifetime $\tau$~$\sim$~800ps. We used filters with central wavelength at 850~$\pm$~10~nm (BP850, blue), 880~$\pm$~10~nm (BP880, green), 905~$\pm$~25~nm (BP905, yellow) and 930~$\pm$~10~nm (BP930, red).}
\end{figure}

\newpage

\begin{figure}[h!]
  \centering
    \includegraphics[width=0.9\textwidth]{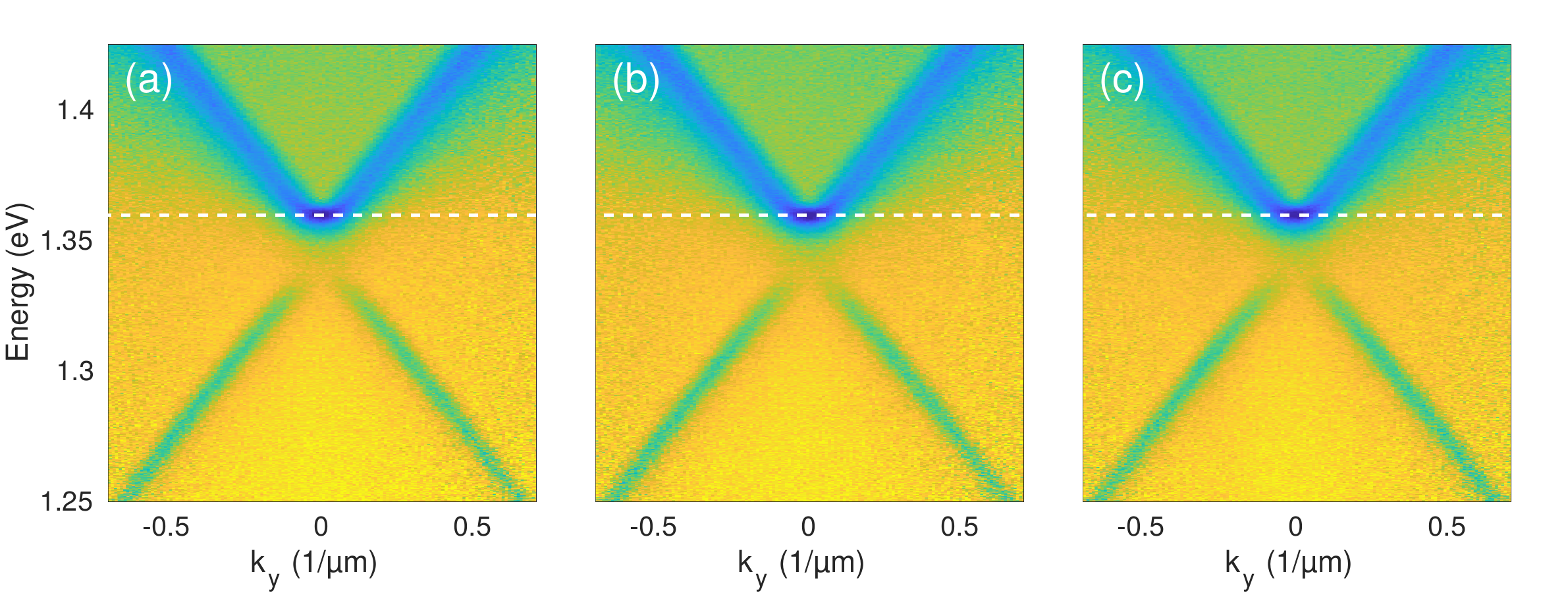}
   \caption*{{\bf Extended Data Figure 12.} The measured dispersions of the p = 600 nm sample measured at the top, middle and bottom part of the array (a, b and c, respectively). The band-edge energy (marked as dashed white line) of the sample remains constant (with accuracy of four digits in eV) in all locations, implying that also the periodicity is constant across the array.}
\end{figure} 

\newpage

\begin{figure}[h!]
  \centering
    \includegraphics[width=0.64\textwidth]{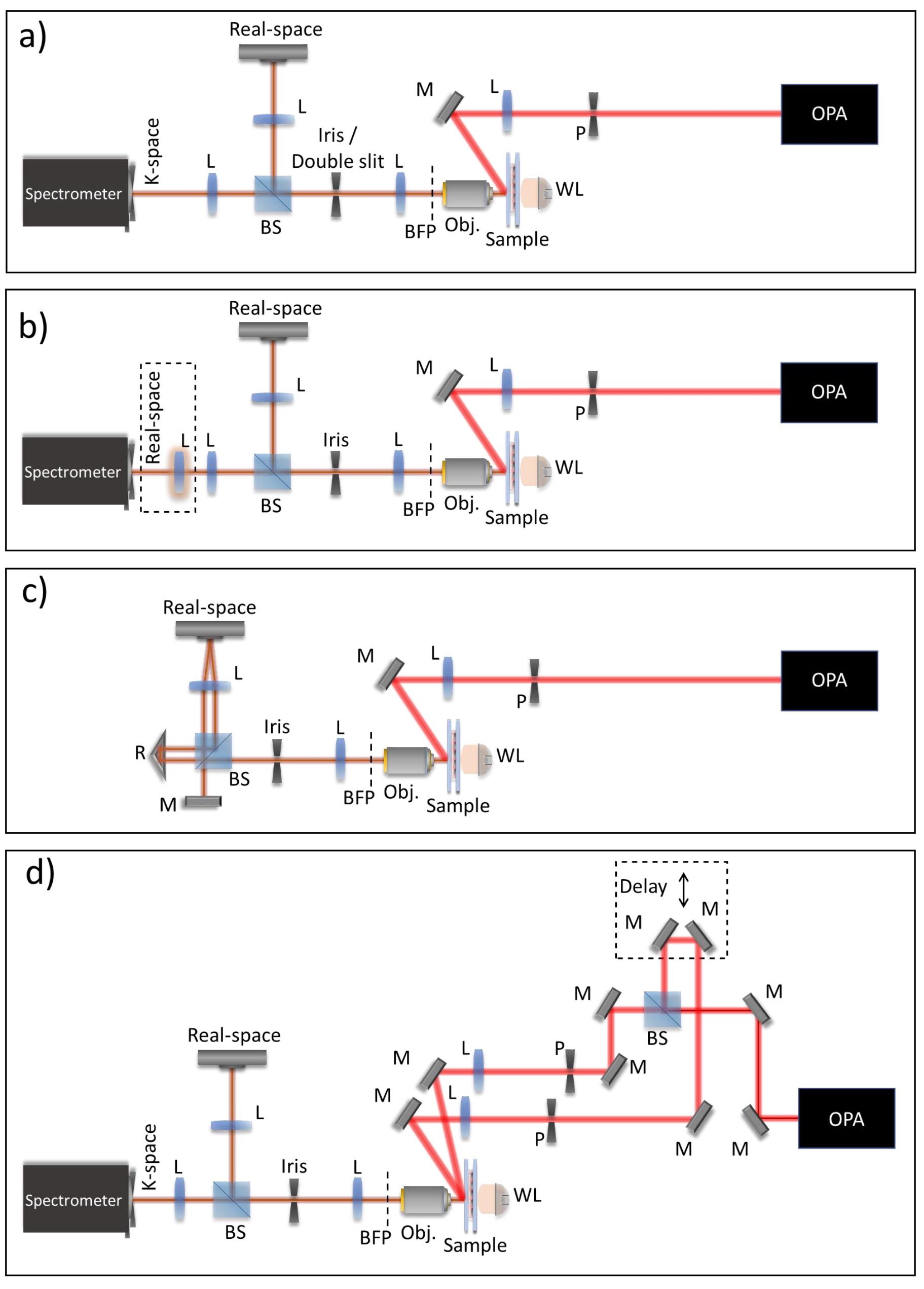} 
   \caption*{{\bf Extended Data Figure 13.} Schematics of the four different measurement schemes that were used. (a) Angle and energy-resolved measurements. Transmission data was obtained by using halogen white light (WL) as a source. For luminescence measurements, the output of an optical parametric amplifier (OPA) was directed through a pinhole (P) and focused to the sample via a lens (L) and a mirror (M). The angle and energy-resolved spectra were obtained by focusing the back focal plane of the imaging objective (Obj.) to the entrance slit of the spectrometer equipped with a two-dimensional CCD. An iris was used to spatially filter the light emerging from the sample. For the double slit experiment, the iris was replaced by a double slit. 
(b) To measure the energy and $y$-position resolved data, an additional lens was used in order to focus the real space image of the sample to the entrance slit of the spectrometer.
(c) To measure the spatial coherence of the BEC on the sample, the spectrometer was replaced by a Michelson interferometer. The beam splitter (BS) was rotated by 90 degrees. One output arm of the beam splitter was directed to a hollow roof L-shaped mirror in order to invert $y$ coordinate to $-y$ upon reflection. The other output arm was reflected back from a regular mirror mounted on a delay stage. Both beams were then directed through the beam splitter. A lens was used to focus both real space images (the inverted and the original one) to a CCD camera.
(d) Pump-probe setup configuration. While the collection side remains the same as a), the excitation side was modified to accommodate a delay line. A same-energy, but highly attenuated, pulse was temporally delayed with respect to the strong pump pulse. Note that to investigate BEC dynamics in our system, we positioned the pump on the edge of the array and the probe at the position where the BEC appears, about 200 $\mathrm{\mu m}$ (center-to-center distance) from each other.}
\end{figure}

\newpage

\begin{figure}[h!]
  \centering
    \includegraphics[width=0.6\textwidth]{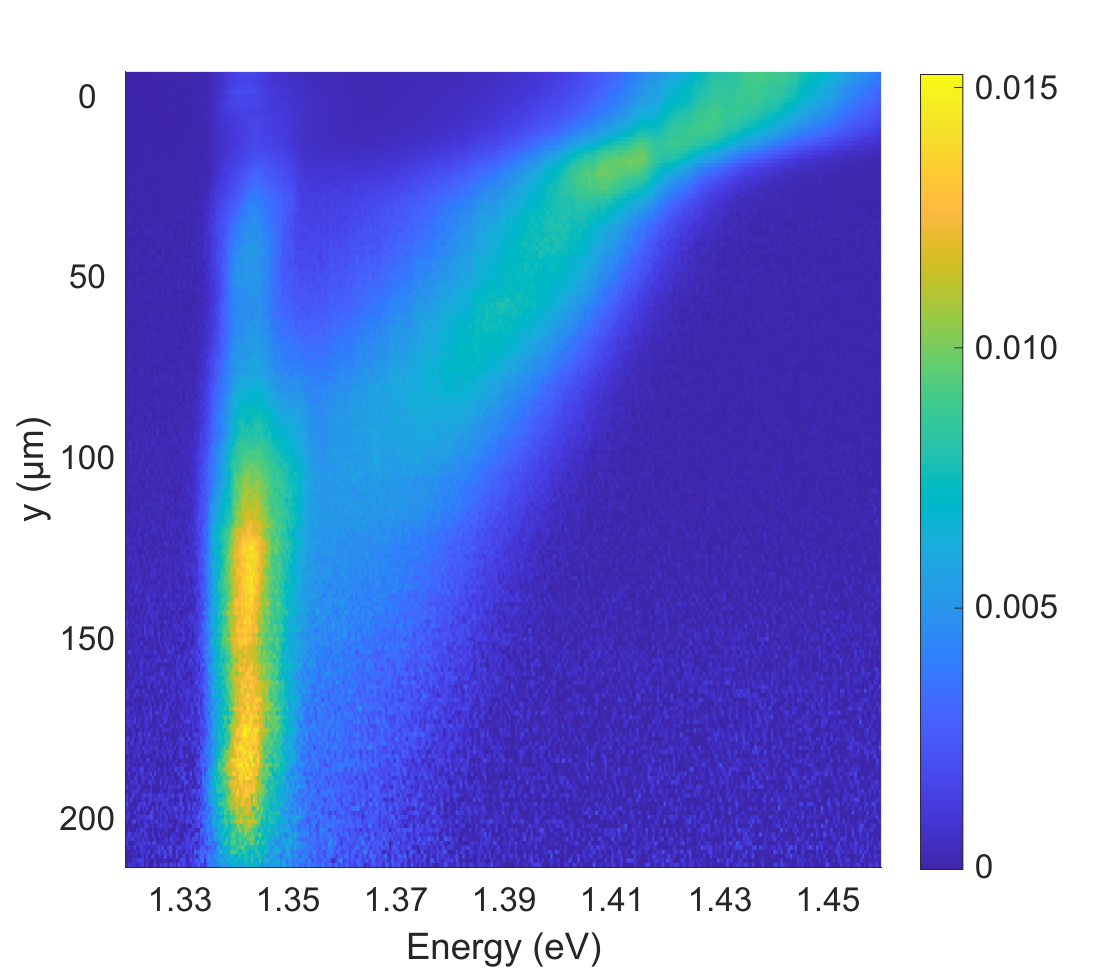}
   \caption*{{\bf Extended Data Figure 14.} A BEC sample ($p = 600$ nm) measured with a broader energy range of the spectrometer. The red-shifting population is clearly visible at higher energies, which then cumulates to the band edge forming a BEC at 1.34 eV. The broader energy range limits the spectral resolution which makes the BEC linewidth at the band edge appear larger than in the figures of the main text.}
\end{figure}

\newpage

\section*{Supplementary Information}

\section*{Rate equations}
Our theoretical rate equation model is based on the approach introduced in \cite{Kirtsupp13,Kirtonsupp15} where a microscopic quantum model was developed to describe the dynamics of photon condensates in cavities. Our system consists of discretized SLR modes which are coupled with $N$ dye molecules. The model assumes that each molecule is a two-level system coupled to a harmonic oscillator that describes the rovibrational degrees of freedom within a molecule. For such a system the microscopic Hamiltonian is of the form
\begin{align}
\label{totalhamiltonian}
H_{\textrm{system}} =  \sum_\textbf{k} \hbar \omega_\textbf{k} a_\textbf{k}^\dag a_\textbf{k} + \sum_{i=1}^N \left( \frac{\hbar \omega_m}{2}\sigma^z_i + \hbar \Omega b_i^\dag b_i - \hbar \Omega \sigma^z_i \sqrt{X}(b_i^\dag + b_i) \right) 
+ \sum_\textbf{k} \sum_i^N \left( g_i a_\textbf{k} \sigma^+_i  + g_i^* a^\dag_\textbf{k} \sigma^-_i  \right).
\end{align}
Here $a_\textbf{k}^\dag$ is the bosonic creation operator of the SLR mode of momentum $\textbf{k}$, $\sigma_i^z$ and $ \sigma_i^\pm$ are the Pauli operators describing the two-level structure of the $i$th molecule and $b_i^\dag$ is the bosonic creation operator corresponding to the vibrational ladder of the $i$th molecule. Furthermore, $\omega_\textbf{k}$ is the energy of the SLR mode of momentum $\textbf{k}$, $\Omega$ is the vibrational energy of the molecules, $g_i$ describes the coupling strength between the $i$th molecule and the SLR modes, $\omega_m$ is the energy of the two-level systems  and $X$ describes the spectral broadening of the molecules due to the vibrational degrees of freedom. In Eq. \eqref{totalhamiltonian} we have assumed that electric field profiles of SLR modes are uniform which is an approximation. In plasmonic lattices the electric fields can have rather complicated profiles and quantizing such fields is extremely challenging.  For details of deriving the above Hamiltonian, see for instance \cite{Julksupp15,Meyssupp07,Mukasupp95,Rodesupp12,Egorsupp08}.

Our system is a highly open quantum system as the SLR modes are extremely lossy, the molecules are externally pumped by a laser beam, and the rovibrational degrees of freedom have small relaxation times. We treat our system by writing a master equation for the time derivative of the reduced density matrix for the SLR modes, $\frac{d \rho}{dt}$, and deploying the standard Born-Markovian approximation for the pump, SLR losses and rovibrational states \cite{Kirtsupp13,Kirtonsupp15,Julksupp15}. With $\frac{d \rho}{dt}$ we can solve the rate of change of the SLR populations $n_\textbf{k} = \langle a_\textbf{k}^\dag a_\textbf{k} \rangle$ by writing $\frac{d}{dt}\langle a_\textbf{k}^\dag a_\textbf{k} \rangle = \textrm{Tr}[a_\textbf{k}^\dag a_\textbf{k} \frac{d \rho}{dt}]$ and the fraction of excited molecules $p_e = \langle \sigma_i^+ \sigma_i^- \rangle$ by having $\frac{d}{dt}\langle \sigma_i^+ \sigma_i^- \rangle = \textrm{Tr}[\sigma_i^+ \sigma_i^- \frac{d \rho}{dt}]$. The rate equations used in this work become:
\begin{align}
\label{rateeqs}
&\frac{d}{dt}n_\textbf{k} = -\kappa(\textbf{k}) n_\textbf{k} + 2N g^2 \left[ \Gamma^+(\textbf{k})(1+ n_\textbf{k})p_e - \Gamma^-(\textbf{k}) n_\textbf{k}(1- p_e) \right] \nonumber \\
&\frac{d}{dt} p_e = P(t)(1-p_e) - \gamma_m p_e + \sum_\textbf{k}2g^2 \left[- \Gamma^+(\textbf{k})(1+n_\textbf{k}) p_e + \Gamma^-(\textbf{k}) n_\textbf{k} (1-p_e)  \right].
\end{align}
Here $\kappa(\textbf{k})$ is the decay rate of the SLR mode of momentum $\textbf{k}$, $\Gamma^\pm(\textbf{k})$ are the molecular emission ($+$) and the absorption ($-$) coefficients for a given SLR mode with momentum $\textbf{k}$, $P(t)$ is
time-dependent pump power and $\gamma_m$ describes the spontaneous non-radiative decay of the molecular two-level systems from the upper energy level to the ground state. One can derive microscopic expressions for $\Gamma^\pm(\textbf{k})$ but in this work we used real emission and absorption profiles of the molecule.

The derivation of Eqs. \eqref{rateeqs} requires multiple approximations. First, we assumed that the correlator term $\langle a_\textbf{k}^\dag a_\textbf{k}\sigma_i^+ \sigma_i^- \rangle$ can be approximated as $ \langle a_\textbf{k}^\dag a_\textbf{k}\sigma_i^+ \sigma_i^- \rangle \approx \langle a_\textbf{k}^\dag a_\textbf{k} \rangle \langle \sigma_i^+ \sigma_i^- \rangle$. This is likely a valid approximation as we are considering the weak-coupling regime. The approximation makes it possible to describe the system completely with Eqs. \eqref{rateeqs} without a need to write extra rate equations for $\langle a_\textbf{k}^\dag a_\textbf{k}\sigma_i^+ \sigma_i^- \rangle$ and other higher-order correlators. 

Furthermore, we  assumed that all the molecules behave similarly. This can be justified by recalling that all the molecules interacting with plasmonic modes are in the proximity of the nanoparticles and thus the volume enclosing the active molecules is small. Therefore we assume that the population inversion $p_e$ and the coupling coefficient $g_i=g$ are the same for all molecules which means we need only one rate equation to describe the dynamics of all of them.

\section*{Spatial coherence measurement}
To study the emergence of spatial coherence, we imaged the condensate by a Michelson interferometer with one of its arms replaced by a hollow roof L-shaped mirror which inverts the image along the $y$-axis. In this configuration we measure the phase coherence between points at $y$ and $-y$. The intensity of the interference pattern between two plane waves is given by~\cite{OpticalPhyssupp}
\begin{align}
\label{Intensity}
I(\boldsymbol{r},\boldsymbol{r}')=A/2 \langle \boldsymbol{E}^*(\boldsymbol{r}) \boldsymbol{E}( \boldsymbol{r}') \rangle = A/2 \langle|\boldsymbol{E}_1|^2 + |\boldsymbol{E}_2|^2 + 2\boldsymbol{E}_1\boldsymbol{E}_2 \cos{\theta} \rangle
\end{align}
where $A$ is a constant related to the speed of light and $\boldsymbol{E}_1= E_1 \textrm{exp} (i[\boldsymbol{k}_1\boldsymbol{r}-\omega t+\phi_1])$ and $\boldsymbol{E}_2= E_2 \textrm{exp} (i[\boldsymbol{k}_2\boldsymbol{r}-\omega t+\phi_2])$ are the electric fields from the arms of the Michelson interferometer and $\boldsymbol{E}_1\boldsymbol{E}_2  \cos{\theta}$ is the interference term responsible for the fringe spacing, orientation and initial phase. The first-order
correlation function, describing the degree of spatial coherence, between the points $\boldsymbol{r}$ and $\boldsymbol{r}'$ is
\begin{align}
\label{contrast}
g^{(1)}(\boldsymbol{r}, \boldsymbol{r}') = \frac{\langle E^*(\boldsymbol{r}) E(\boldsymbol{r}') \rangle }{\langle E^*(\boldsymbol{r}) \rangle \langle \langle E(\boldsymbol{r}') \rangle}.
\end{align}
When the mirror image is taken centrosymmetrically (in our case $\boldsymbol{y}' =
-\boldsymbol{y}$), the correlation function is obtained directly from the visibility of the interference fringes, as
\begin{align}
\label{contrast}
C(\boldsymbol{y}, -\boldsymbol{y}) &= \frac{I_\textrm{max}-I_\textrm{min}}{I_\textrm{max}+I_\textrm{min}} = \frac{2 \sqrt[]{I(\boldsymbol{y})I(-\boldsymbol{y})}}{I(\boldsymbol{y})+I(-\boldsymbol{y})} g^{(1)}(\boldsymbol{y}, -\boldsymbol{y}).
\end{align}
Here, $I(\boldsymbol{r})$ is the light intensity and $I_\mathrm{max}$ and $I_\mathrm{min}$ are the maximum and minimum intensities of the fringes, respectively. Note that in our case the contrast is expected to decrease for increasing $y$ due to the decay ($I(\boldsymbol{y})<I(-\boldsymbol{y})$) of the SLR population. 

\section*{Estimating photon densities}

\subsection*{Critical photon density for BEC}
We calculate the critical photon density for BEC in a finite 2D system~\cite{Khalasupp65,Ketterlesupp96,deng_exciton-polariton_supp2010,PethickSmithsupp}. Assuming the SLR mode dispersion to be parabolic around the band bottom, 
\begin{align}
\label{dispersion}
E(\boldsymbol{k}) &= E_0 + \frac{\hbar ^2 k^2}{2 m_{eff}},
\end{align}
we obtain an estimate for the effective mass $m_{eff}$. Here, $\hbar$ is the reduced Planck's constant $h/2\pi$ and $k$ the momentum. Fitting Eq. \eqref{dispersion} to the band bottom gives $m_{eff} = 1.7 ... $ $2.6 \times 10^{-37}$ kg. The values $1.7 \times 10^{-37}$ kg and $2.6 \times 10^{-37}$ kg are obtained by two equally reasonable fits where different range of wavevectors around k = 0 are selected as the band bottom. With the standard 95 \% confidence level, the lower and upper error limits (standard 95 \% confidence level) for these values are $1.3 \times 10^{-37}$ and $4.3 \times 10^{-37}$, respectively. The respective 99 \% confidence bounds for the values are $1.2 \times 10^{-37}$ and $5.8 \times 10^{-37}$. Critical density ($n_c=N/V$) is given by
\begin{align}
\label{criticaln}
n_c &= \frac{2}{\lambda_T^2} \textrm{ln}(\frac{L}{\lambda_T}),
\end{align}
where $L$ is the size of the 2D box. We take the measured coherence length of the SLR modes in absence of molecules ($45 \mathrm{\mu m}$) as the box width. Thermal de Broglie wavelength is defined as
\begin{align}
\label{debroglie}
\lambda_T(T) &= \sqrt{\frac{2 \pi \hbar^2}{m k_B T}},
\end{align}
where $k_B$ is the Boltzmann constant. Now, using Eq. \eqref{criticaln} and the effective mass, the critical photon density is between $n_c = 3.0 ... 5.3 \times 10^{10}$ 1/m$^2$. Within 95 \% confidence bounds for the effective mass, the critical photon density is between $n_c = 2.1 ... 10.0 \times 10^{10}$ 1/m$^2$. With the 99 \% confidence bounds, the critical photon density is between $n_c = 1.8 ... 14.6 \times 10^{10}$ 1/m$^2$.
Repeating such calculation for 2D linear and 1D parabolic dispersions gives the same order of magnitude, $n_c \simeq 10^{10}$ 1/m$^2$, while 1D linear gives $n_c \simeq 10^{9}$ 1/m$^2$ (finite size is required in the 1D linear/parabolic cases but not in 2D linear one). Therefore the estimate $n_c \simeq 10^{10}$ 1/m$^2$ is reasonable for our real dispersion which is a combination of linear and flatter regions, and is anisotropic. One has to also bear in mind that these are equilibrium values while our system is open-dissipative and can at best reach a quasi-equilibrium.

\subsection*{Lower bound for the photon density in the experiment}

As described in Methods, the sample luminescence was measured with the 2D charge coupled device (CCD) camera. The CCD intensity reading was calibrated against a 633 nm HeNe laser, taking into account the wavelength-dependent quantum yield of the CCD. The emission was collected through the spectrometer entrance slit whose width was 3/4 of the array width on the sample.

By summing the collected intensity over the part of the sample where the condensate is observed, we get the total number of counts in the CCD to be around $4.98 \times 10^{4}$. Taking into account the overall efficiency of the CCD (limited by instrument response, quantum yield, etc.), we measured that each CCD count requires 20 photons. The luminescence is measured only from one side of the sample, and we assume the sample to radiate equally to both sides, so the actual photon number in the system is twice the observed. Therefore, the lower limit for the total number of photons in the BEC is around $1.99 \times 10^{6}$.

As the integration time in the luminescence measurements was 300 ms, and the laser pulse frequency is 1 kHz, we can deduce the number of photons in the BEC per pulse to be $6.63 \times 10^{3}$ . Dividing this by the area from where the emission was collected, 75 $\times$ 100 $(\mathrm{\mu m} ) ^2 $, the lower bound for the number density of photons in the BEC is around $8.8 \times 10^{11}$ 1/m$^2$. 

\bibliographystyle{apsrev4-1}

\end{document}